\documentclass[11pt,preprint]{aastex}

\newcommand{\simgt}{\ga}
\newcommand{\simlt}{\lower.5ex\hbox{$\; \buildrel < \over \sim \;$}}

\shorttitle{Infrared Spectrum of Zodiacal Light}
\shortauthors{Reach {\it et al.}}

\begin{document}

\title{The Mid-Infrared Spectrum of the Zodiacal and Exozodiacal Light}

\author{William T. Reach and Patrick Morris}

\affil{SIRTF Science Center/Infrared Processing and Analysis Center, 
California Institute of Technology,
Pasadena, CA 91125}

\email{reach@ipac.caltech.edu}

\author{Fran\c cois Boulanger}
\affil{Institut d'Astrophysique Spatiale, 91405 Orsay cedex, France}

\author{Koryo Okumura}
\affil{CEA/DSM/DAPNIA Service d'Astrophysique, F-91191 Gif-sur-Yvette, France}

\def\extra{
\slugcomment{Manuscript Pages: 78, Tables: 5, Figures: 15}

\noindent {\bf Proposed running head:} Infrared Spectrum of Zodiacal Light
\bigskip

\noindent {\bf Editorial correspondence to:} 

Dr. William T. Reach

SIRTF Science Center

MS 220-6

Caltech

Pasadena, CA 91125

Phone: 626-395-8565

Fax: 626-432-7484

E-mail: reach\@ipac.caltech.edu
\clearpage
}

\begin{abstract}

The zodiacal light is the dominant source of the mid-infrared sky brightness
seen from Earth, and exozodiacal light is the dominant emission from 
planetary and debris systems around other stars.
We observed the zodiacal light spectrum with 
the mid-infrared camera ISOCAM over the wavelength range 5--16 $\mu$m
and a wide range of orientations relative to the Sun (solar elongations
68$^\circ$--113$^{\circ}$) and the ecliptic (plane to pole). 
The temperature in the ecliptic ranged from 269 K at solar 
elongation $68^\circ$ to 244 K at $113^\circ$, and
the polar temperature, characteristic of dust 1 AU from the Sun, is 274 K.
The observed temperature is exactly as expected for
large ($> 10 \mu$m radius), low-albedo ($< 0.08$), rapidly-rotating,
grey particles 1 AU from the Sun.
Smaller particles ($< 10$ $\mu$m radius) radiate inefficiently in the infrared
and are warmer than observed.
We present theoretical models for a wide range 
of particle size distributions and compositions;
it is evident that the zodiacal
light is produced by particles in the 10--100 $\mu$m radius range.
In addition to the continuum, we
detect a weak excess in the 9--11 $\mu$m range, with an amplitude
of 6\% of the continuum. The shape of the feature can be matched
by a mixture of silicates: amorphous forsterite/olivine provides most of
the continuum and some of the 9--11 $\mu$m silicate feature, dirty crystalline olivine
provides the red wing of the silicate feature (and a bump at 11.35 $\mu$m),
and a hydrous silicate (montmorillonite) provides the blue wing of
the silicate feature. The presence of hydrous silicate suggests the parent
bodies of those particles were formed in the inner solar nebula.
Large particles dominate the size distribution,
but at least some small particles (radii $\sim 1$ $\mu$m) 
are required to produce the silicate emission feature. The strength
of the feature may vary spatially, with the strongest features being
at the lowest solar elongations as well as at high ecliptic latitudes;
if confirmed, this would imply that the dust properties change such
that dust further from the Sun has a weaker silicate feature.

To compare the properties of zodiacal dust to dust around other main
sequence stars, we reanalyzed the exozodiacal light spectrum for $\beta$ Pic
to derive the shape of its silicate feature.
The zodiacal and exozodiacal spectra are very different.
The exozodiacal spectra are dominated by cold dust, with emission
peaking in the far-infrared, while the zodiacal spectrum peaks
around 20 $\mu$m. 
We removed the debris disk continuum from the spectra by fitting a
blackbody with a different temperature for each aperture (ranging
from $3.7''$ to $27''$); the resulting silicate spectra for $\beta$ Pic
are identical for all apertures, indicating that the silicate feature arises
close to the star.
The shape of the silicate feature from $\beta$ Pic is nearly identical to that
derived from the {\it ISO} spectrum of 51 Oph;
both exozodiacal features are very different from that of the zodiacal light.
The exozodiacal features are roughly triangular, peaking at 10.3 $\mu$m,
while the zodiacal feature is more boxy. 

\noindent{\bf Keywords}: zodiacal light, infrared observations, 
interplanetary dust

\end{abstract}


\clearpage

\section{Introduction}

The spectrum of the zodiacal light is a key for understanding the origin 
of Solar System dust disk and its connection to dust disks around other
planetary or planet-forming stellar systems. 
Theoretically, we expect a
distinction between spectra of disks around older and younger disks,
as the originally small interstellar dust particles 
combine to form planetesimals, comets, and asteroids \citep{beckwith}.
Young stars have disks dominated by small particles, which 
produce a silicate feature diagnostic of particle chemistry.
Toward the well-studied star $\beta$ Pic, the 10 $\mu$m 
silicate feature was detected \citep{telknack} and its substructure measured
\citep{knacke}. Over the entire infrared spectrum of
the well-observed debris disk around the star HD 14257, several silicate
features were detected, and these features are by far the
dominant structures in the spectra \citep{malfait}. 
For a sample of AeBe stars, silicates
are generally present but with significant variations form star
to star \citep{muess}.
The asteroids and comets of a mature planetary system produce 
dust via ice sublimation and mutual collisions. While small
particles are ejected from the solar system by radiation pressure,
larger particles 
must be continually replenished to balance losses due to 
Poynting-Robertson drag and fragmentation due to 
mutual collisions \citep{grun85}.
The differences in particles from star to star may be
reflected in the variety of glass
 with embedded metal and sulfide (GEMS)
inclusions in meteorites, some of which look
like astronomical silicates characteristic of the diffuse
interstellar medium, while others are more crystalline,
as seen in the spectra of disks and new comets \citep{bradley99}.
 Crystalline silicates like SiC are present in meteorites, and they are
 thought to originate in the solar system since crystalline silicates 
 are not observed in the ISM, possibly destroyed by supernovae shocks during
 transport from the envelopes of evolved stars \citep{demyk}.
 \citet{molster} suggest that large crystalline silicate grains containing iron (Fe)
 may have been amorphous grains annealed close to the sun, then transported
 outwards by turbulence in the early solar disk, while some Fe-poor crystalline
 silicates in small grains may have condensed from the gas phase in the inner
 disk, then transported by radiation pressure {\it over} the solar disk to distances
 beyond Jupiter.  
Interplanetary dust particles captured in the
Earth's atmosphere are predominantly silicate 
minerals with prominent 10 $\mu$m features in their absorption
spectra \citep{sandfordwalker}.
The spectrum of the Solar System dust is a `laboratory'
measurement, because we actually know some of the properties
of interplanetary dust by {\it in situ} observation.
The zodiacal spectrum is potentially a key to 
discriminating between old and young disks around other stars.

The particles that produce the zodiacal light originate from both
asteroids and comets. The mix of cometary and asteroidal dust 
probably varies, depending on particle size, composition,
and orbital distribution.
New comets produce crystalline particles with spectra similar
to young proto-planetary dust disks \citep{crovisier,hayward}.
Analyzing the dynamics of particles produced by
the `mature' short-period comet 2P/Encke revealed that it
produces only mm-sized and larger particles \citep{reachencke}.
The broad-band infrared spectrum of the zodiacal light
over 12--100 $\mu$m is matched by a model with particles
composed of silicates and the same size distribution 
measured {\it in situ} by Earth-orbiting satellites
\citep{grun85,reach88}.
In addition to size and composition, orbits may
distinguish asteroidal and cometary particles.
As asteroids are relatively more confined to the ecliptic
than comets, we might expect a difference between       
the mineralogy of the dust between observations in the ecliptic
plane and toward the ecliptic pole.

The goals of this paper are to determine whether there is a silicate
feature in the zodiacal light and to determine whether there are
spatial variations in the spectrum of the zodiacal light.
Previous work indicated there may be some structure in the mid-infrared
zodiacal light spectrum, but at a level of no more than 15\% of the continuum.
This result was obtained using the spectrum of the zodiacal light 
from 5--16 $\mu$m measured toward one line of sight 
as part of the performance verification phase of the {\it Infrared Space
Observatory} \citep{reach96}.
The results showed a hint of a spectral feature with shape similar 
to that seen from comets \citep{hanner94,crovisier,grun01} and 
the $\beta$ Pic debris disk \citep{knacke,fajardo95}. 
Some evidence for a silicate feature in the zodiacal light was also
found from observations with the Infrared Telescope in Space (IRTS)
\citep{ootsubo98,ootsubothesis}.
It is important not
only to determine whether there is a silicate spectral feature but
also to constrain its shape. Cometary and $\beta$ Pic dust
have already been shown to have a 10 $\mu$m silicate feature with shape
different from that of interstellar dust \citep{dl84}
from which Solar System solids formed. 
Previous observational results for the zodiacal light
were inconclusive because of potential
detector and optical defects that had not yet been characterized,
or because the spectral coverage did not extend beyond the red
edge of the silicate feature (IRTS).
In this paper, we present follow-up {\it ISO}
observations and analysis of deep archival data
that have the optimal sensitivity to the diffuse emission.
The new observations include the ecliptic pole, ecliptic plane, and
mid-latitudes, and we compare the spectra to determine whether
there is evidence for different types of particles overhead as
compared to particles in the ecliptic, both near the Earth's
orbit and exterior to it. Then we compare the zodiacal light
observations to exozodiacal spectra, to search for
similarities and differences between Solar System dust
and dust around other main sequence stars.

\section{Observations}

\subsection{Zodiacal light: Observing strategy\label{sec21}}

The new observations were made using the ISOCAM \citep{cesarsky} instrument
on the {\it Infrared Space Observatory} \citep{kessler}.
They consist of 3 observing sequences with the circular-variable filter
(CVF); two observations were in the ecliptic plane and one was toward the 
ecliptic pole.  
The observations in the ecliptic plane were performed as two-epoch,
fixed-time observations of the same piece of sky, to measure the 
change of the brightness with the changing position of the Sun.  
These observations were scheduled at different times of year such      
that the solar elongations ($\epsilon_i$) on the two dates are as
close as possible to supplemental angles 
($\epsilon_1+\epsilon_2=180^\circ$).  
The widest separation of solar elongations makes it more possible to separate
the contributions from the inner and outer Solar System.
Taking the visibility constraints of {\it ISO}, 
allowing some margin to make it possible to schedule the observations,
and choosing the ecliptic plane observations to be taken as far from
the galactic plane as possible,
the observed lines of sight were toward ecliptic longitude $50.1^\circ$
and latitude $0.3^\circ$ with solar elongations 
$\epsilon_1=68^\circ$ and $\epsilon_2=113^\circ$.
The same astronomical observing template (AOT)               
configuration was used for each spectrum: 10 frames of 5 sec duration 
were taken at each step of the long-wavelength circular variable filter (CVF).  
We used the $12''$ pixel-field-of-view lens in order to maximize the illumination 
on the detector.  Transient gain response is an important limitation of 
background observations like these, and the timescale for transient response is inversely             
proportional to the flux on the detector.  By using the $12''$
pixel-field-of-view lens and 5 sec integration time, the flux is increased      
by a factor of 10 relative to the previous ISOCAM zodiacal light spectrum, which       
allowed us to reach the fainter brightness levels at the ecliptic          
pole.  After each CVF observation, an observation of the same field with        
three filters (LW6, LW7, and LW8) was concatenated.  These observations were
used to confirm that the detectors were stabilized during the
CVF observations. The observing conditions are summarized in ~\ref{isolog}.

\begin{table}[th]
\caption{\it ISOCAM Observing Log}\label{isolog} 
\begin{center}
\begin{tabular}{llrrrrrrr} \hline\hline
& & & \multicolumn{2}{c}{Ecliptic (J2000)} && \multicolumn{3}{c}{DIRBE} \\
\cline{4-5} \cline{7-9} 
\multicolumn{1}{l}{Project} & 
\multicolumn{1}{r}{Day} & \multicolumn{1}{r}{$\epsilon$} &
\multicolumn{1}{c}{$\lambda$} & \multicolumn{1}{c}{$\beta$} & &
\multicolumn{1}{c}{4.9 $\mu$m}  & \multicolumn{1}{c}{12 $\mu$m}  &
\multicolumn{1}{c}{25 $\mu$m} \\
\cline{1-9}
WREACH.ZODYSP & 200 &  68.9 &  50.12 &   0.30 &&   2.03 &  64.48 &  79.42 \\
WREACH.ZODYSP & 249 & 113.9 &  50.12 &   0.30 &&   0.50 &  25.82 &  37.28 \\
WREACH.ZODYSP & 118 &  89.6 &   0.48 &  89.48 &&   0.40 &  13.06 &  15.50 \\
       LUTZ00 & 334 &  78.0 & 324.70 & -20.22 &&   1.09 &  35.08 &  41.89 \\
       LUTZ01 & 341 &  81.4 & 257.81 &  81.41 &&   0.48 &  13.83 &  15.59 \\
       LUTZ02 & 344 &  74.0 & 312.29 & -61.49 &&   0.63 &  16.91 &  18.76 \\
       LUTZ03 & 344 &  91.5 & 354.13 & -76.94 &&   0.43 &  13.40 &  15.53 \\
       LUTZ04 & 345 & 107.9 &  49.32 & -69.40 &&   0.35 &  12.55 &  15.14 \\
       LUTZ05 & 347 &  88.9 & 349.60 & -18.80 &&   0.84 &  30.85 &  38.51 \\
       LUTZ06 & 348 &  84.5 & 345.57 & -25.85 &&   0.82 &  27.38 &  32.97 \\
       LUTZ07 & 352 & 106.8 &  17.88 & -39.87 &&   0.45 &  16.91 &  21.02 \\
       LUTZ08 &	3 & 104.6 &  30.58 & -36.30 &&   0.47 &  17.60 &  21.91 \\
       LUTZ09 &	4 &  74.8 & 355.26 & -32.59 &&   0.88 &  25.28 &  28.96 \\
       LUTZ10 &	9 &  68.5 & 353.66 & -29.10 &&   1.12 &  29.65 &  33.30 \\
       LUTZ11 &	9 &  72.6 &   0.75 & -10.07 &&   1.56 &  49.38 &  59.69 \\
       LUTZ12 &  19 & 115.6 &  55.89 & -19.27 &&   0.47 &  20.89 &  27.77 \\
       LUTZ13 &  25 &  88.0 &  32.09 & -42.15 &&   0.54 &  17.54 &  20.71 \\
       LUTZ14 &  36 &  75.9 &  31.53 & -10.19 &&   1.31 &  43.87 &  53.80 \\
       LUTZ15 &  40 &  99.4 &  59.32 &   3.42 &&   0.78 &  33.92 &  45.38 \\
       LUTZ16 &  92 &  99.6 & 270.87 &  30.54 &&   0.52 &  20.25 &  25.46 \\
\hline
\end{tabular}
\end{center}
\end{table}

The ISO Short Wavelength Spectrometer \citep{degraauw} was also used in calibration 
and discretionary time to
obtain deep AOT6 observations ($\lambda/\Delta\lambda \sim 3000$, multiple 
scans at the longest integration times) for the purpose of detecting the 
thermal continuum  and possible spectral features from zodiacal dust 
grains.  The observations were obtained in revolutions 461, 769, and 836 at 
positions which corresponded to solar elongations of maximum viewable zodiacal 
brightnesses.  Emission was detected in the 7 to 12 $\mu$m range, 
to within the 1-$\sigma$ detection threshold for the SWS Band 2 detectors 
of $\sim$50 MJy sr$^{-1}$, but the sensitivity was insufficient to 
allow a search for spectral features.  No corrections to the SWS
relative spectral response functions for background emission were ever
applied, or needed as an outcome of these measurements. The observed
brightness was consistent with that predicted by the COBE/DIRBE zodiacal
light model \citep{kelsall} described in more detail in \S\ref{secdark}.

\subsection{Zodiacal light: Archival data}

To supplement the dedicated observations described above, we searched for 
archival {\it ISO} data suitable for deriving zodiacal light spectra. We 
concentrated only on ISOCAM CVF data, which is by far the most sensitive 
data for this task. We attempted to derive a zodiacal light spectrum from
the deepest spectrometer (SWS) data, but the spectra were inadequate 
because the zodiacal light was too faint in comparison to the dark 
current.
One ISOCAM program in particular, DLUTZ.ZZULIRG, was ideal for use because
(1) their intended targets were faint ($< 70$ mJy);
(2) the targets were expected to be point-like; 
(3) the observations are very deep (over 2 hours per pointing);
and (4) the spectra were taken {\it aller-retour}, i.e. rotating
the CVF toward increasing wavelengths and then back toward decreasing
wavelengths. The scientific results on the intended targets
were presented by \citet{tran01}.
Table~\ref{isolog} shows the observing parameters for these observations;
we will refer to specific archival spectra by the sequential numbers 
listed after `LUTZ' in this table.

\subsection{Zodiacal light: Data reduction}


\subsubsection{Dark current and zero-point}\label{secdark}
The observations were analyzed using routines developed at the {\it Institut
d'Astrophysique Spatiale} (IAS) as part of the pre-launch calibration of
ISOCAM and the performance verification (PV) phase of the {\it ISO} mission.
For each image, we used a dark-current image obtained during the PV phase.
To correct for the drift of the dark current between PV phase
and the epoch of observation, a correction was obtained using the 
portion of ISOCAM that is not illuminated with the 12" lens and Fabry
mirror combination. We avoided the region near the outer edge of the 
Fabry mirror, which was contaminated by stray light.
The uncontaminated `skirt' region was correlated pixel-for-pixel
with the library dark. The library dark was offset and scaled using
the zero-point and slope of this fit, and then it was subtracted from
each image. 

The brightness at the shortest wavelengths was then
then compared to the predicted brightness of the zodiacal light.
Our model is essentially identical to the one derived
by the {\it Cosmic Background Explorer} Diffuse Infrared
Background Experiment (COBE/DIRBE) \citep{kelsall}.
This three-dimensional structure of the interplanetary dust complex includes 
a smooth, tilted dust cloud, which dominates the brightness, as well as anisotropies
due to the Earth's resonant dust ring \citep{dermottring,reachring}
and the asteroidal dust bands \citep{dermottband,reachband}.
The wavelength-dependence is through the Planck-function kernel of
the brightness integral as well as a set of albedoes and emissivities for
each of the 10 DIRBE wavelebands spanning 1.25--240 $\mu$m. 
We have generalized the model so that it predicts the brightness at
an arbitrary wavelength \citep{reachmemo}. We replaced
these values with smooth curves for the wavelength-dependent emissivity 
and albedo. Similarly, the scattering phase-function, which was parameterized
separately for each DIRBE waveband, was replaced by an interpolation between 
the phase functions DIRBE-derived phase functions. At ISOCAM wavelengths,
scattering is only a very minor contributor to the brightness.
Zodiacal model predictions were made for the appropriate celestial coordinates, 
observing date, and wavelength of each CVF observation. 
A small offset was then applied to the ISOCAM data, 
to make the brightness at 5.5--6 $\mu$m match the COBE/DIRBE model.

\subsubsection{Transient response of the detectors}
These observations are at low light levels, where the observed brightness
can be affected by transient gain response and memory effects in the
detectors. We carefully checked
that the detectors were measuring the actual spectrum of the sky as the
CVF was rotated (as opposed to long time-constant transient gain responses).
To this end, the IAS transient correction method \citep{coulais} was applied.
This method is a straightforward calculation that does not involve fitting
to the data and is specified by few parameters.
The transient correction was verified by comparing the spectra taken with
the CVF rotating in both the increasing- and decreasing-wavelength
directions. The transient effect makes the detector output rise more
slowly than the signal input; thus, when the brightness was
increasing, the observed signal is lower than the true signal. 
Figure~\ref{figtransient} shows the results of the transient correction.
The first half of the observation was taken in the decreasing-wavelength,
decreasing-brightness direction, where the transient corrections were
smaller. After correction, the decreasing- and increasing-wavelength
spectra are consistent to within 1--2\% of their average, with the
decreasing-wavelength spectrum consistently brighter. This suggests 
that part of the transient gain response has not been corrected, but there
is relatively little structure as a function of wavelength that could
affect detection of spectral features.
We incorporate the differences between the spectra taken
in each direction into the systematic errors.

\epsscale{0.75}
\plotone{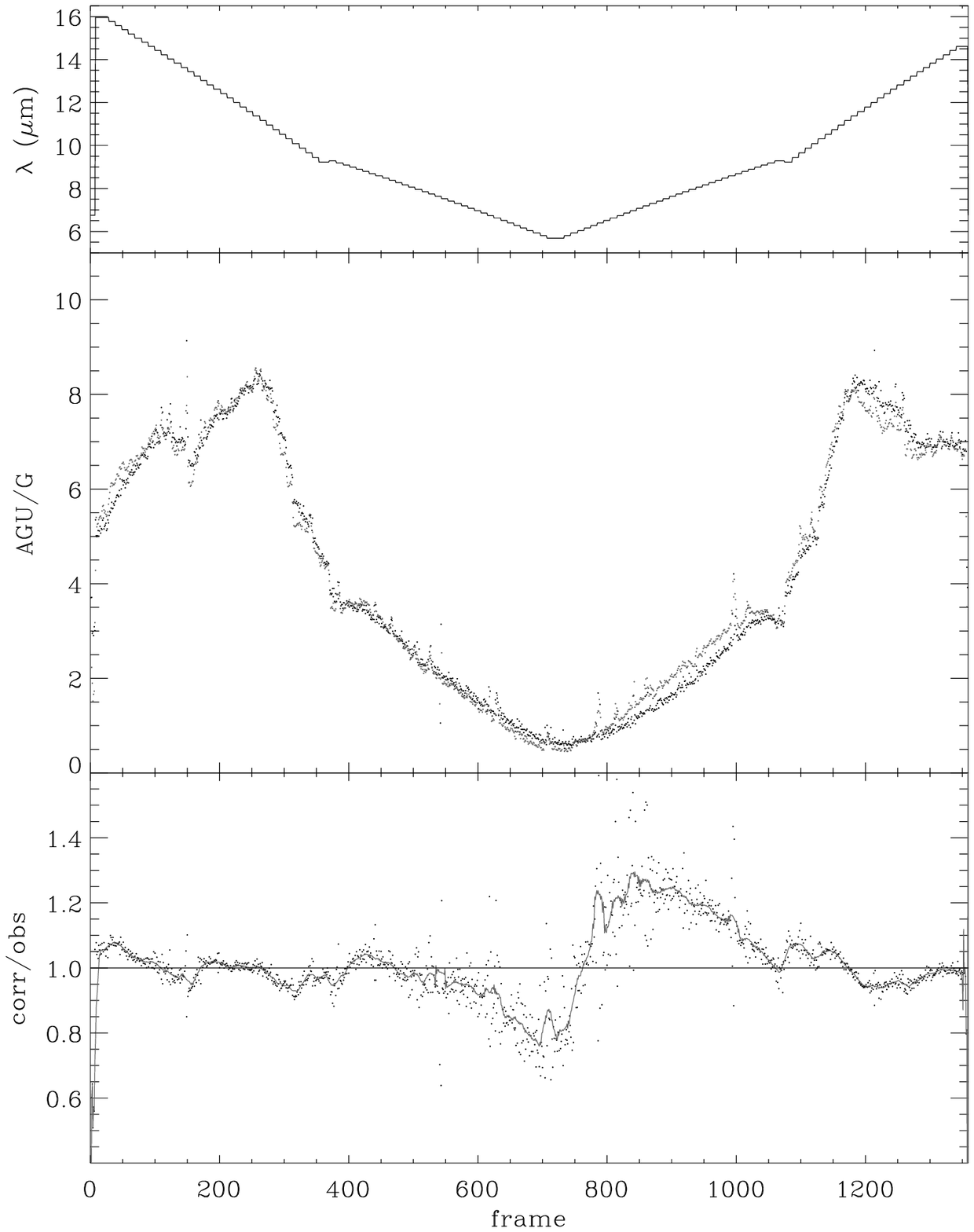}
\epsscale{1}
\figcaption{Illustration of the correction for ISOCAM CVF
transient gain response.
{\protect\it (a)} The top panel shows the wavelength as a function of frame number for
the observation of the north ecliptic pole. The CVF was cycled from
long to short wavelength, then back up. 
{\protect\it (b)} The middle panel shows the brightness (in electrons per second)
averaged over the central part of the each frame. 
One set of dots shows the original data, and another set of dots shows the
transient-corrected data. The detector response
for decreasing signals is faster than that for increasing signals, so the
observed and corrected data are very similar for 
the first half of the observation.
The time constant for transient response is longest for
low brightness levels, so the largest deviations are near the beginning
of the second half of the observation, where the sigla was weak and increasing.
{\protect\it (c)} The bottom panel shows the transient correction factor 
(corrected divided by observed brightness) as a function
of frame number.
\label{figtransient}}
\raggedright

\subsubsection{Stray light removal}
A new and significant correction to the spectra was to remove stray 
light \citep{koryostray}. 
This effect was a significant limitation of our previous attempt to
measure the absolute sky spectrum \citep{reach96}. Absolute measurements
are most prone to stray light because it is exceptionally difficult to
differentiate diffuse stray light from the true sky background. Most
flat-field calibration observations include stray light, which is desired
because the stray light represents a real background to the detector.
For the present problem of separating stray light from the true background,
we used dedicated calibration observations of a point source observed
over the range of CVF wavelengths and at two different locations on the array.
Comparing the background observed through all of the ISOCAM filters,
spanning 4.5--15 $\mu$m, to the brightness observed with the CVF reveals 
systematic differences that we attribute to stray light in the CVF.

A point source observed through the CVF generates `ghost images'
on the array; the ghosts move as the source is moved. 
The ghost images
are due to multiple reflections in the optics, and these reflections will
affect the diffuse emission in a manner similar to the point sources.
The diffuse
stray light amplitude should be equal to the average amplitude of the 
ghost images for point sources located all over the array.
In practice, we only have calibration data for two pointings of the point
source on the array. Thus, we are forced to assume that the wavelength-dependence
of the stray light for uniform illumination is the same as that of the average of
the ghost amplitudes for two pointings on the array.

For both pointings of the calibration point source,
the ghost fluxes were measured and normalized to
the main point source flux. This clearly showed the spectral dependence
of the ghost intensity. We used the average of the ghost flux spectra
for the two pointings of the calibration star to extrapolate the 
ghost intensity of the point source to the zodiacal light photometric
region of the detector; see \citet{blommcvf} for further detail.
ISOCAM was calibrated using the central point-spread-function of stars
(not including the ghost images) so that extended emission
(which includes stray light) must be corrected. 
The stray light amplitude depends on wavelength:
it decreases monotonically from 35\% at 6 $\mu$m to 20\% at 9.3 $\mu$m
near the edge of the short-wavelength CVF; then for the long-wavelength CVF
it jumps to 30\% at 9.5 $\mu$m, decreasing to 6\% at 15.3 $\mu$m.
The stray-light correction is discontinuous across the boundaries between
the two CVFs at 9.5 $\mu$m. This agrees with each of our zodiacal light
observations, which showed a significant discontinuity before 
stray-light correction.
The stray-light correction was validated by comparing to data taken through
9 different ISOCAM filters. The filter data have very coarse spectral 
resolution ($\lambda/\Delta\lambda\sim 5$),
but they do not suffer as much from stray light because the filters are
tilted with respect to the detector. 
To compare the filter and CVF data, we first divided each intensity measurement
by the DIRBE zodiacal light model (\S\ref{secdark}) appropriate for the
observing date, direction, and wavelength. 
Figure~\ref{filter_data} shows the comparison between 
the CVF and filter data \citep{blommcvf}. 
The spectra will be discussed in detail in 
the remainder of this paper, but this figure is shown in advance as a 
calibration validation.
Before correction, there were clear discrepancies
between the CVF and filters. After correction, the two types of observation 
agree well. Both the filters and CVF seem to show systematic deviations,
at the 10\% level, from the DIRBE zodiacal light model.

\plotone{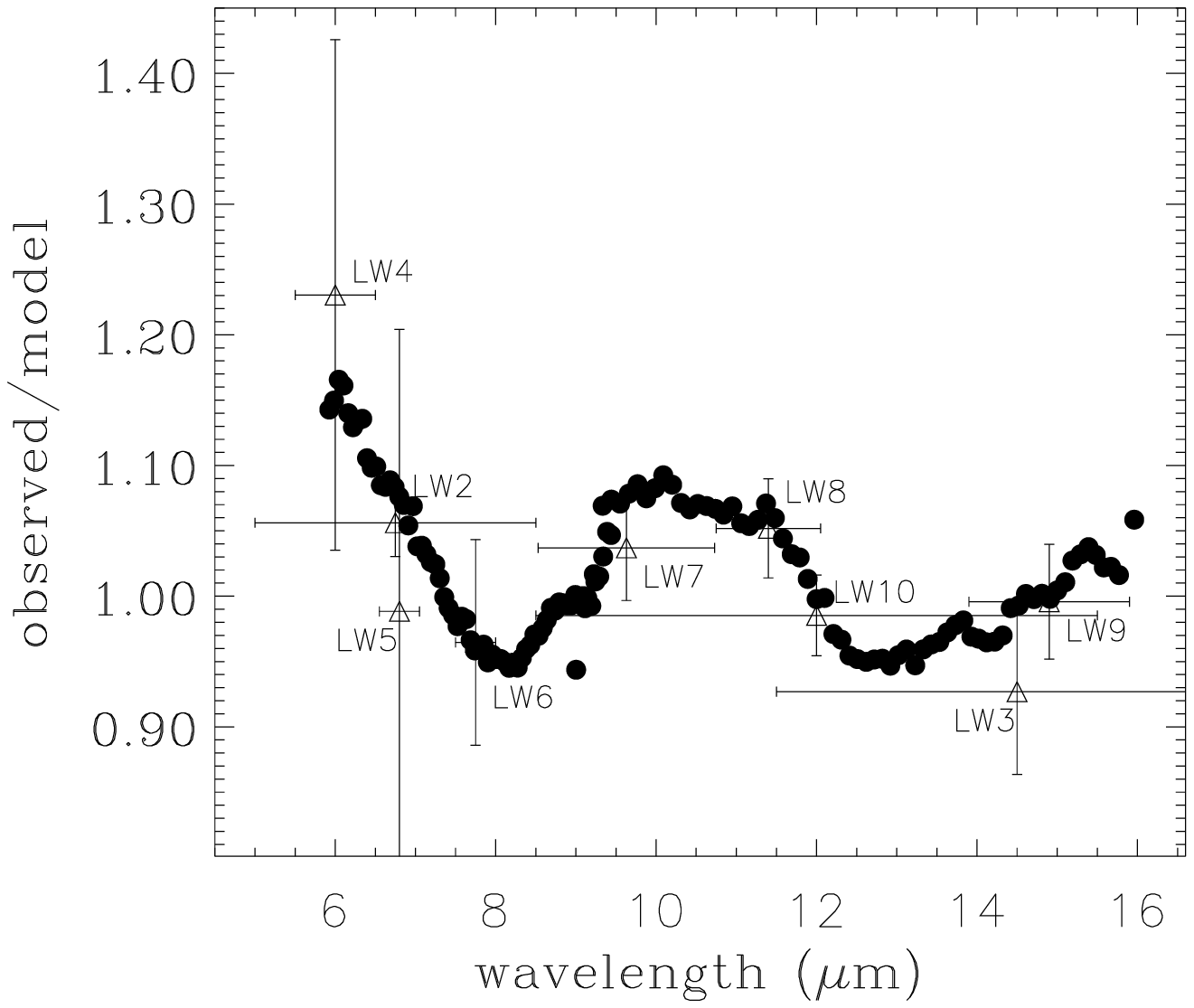}
\figcaption{Comparison between the sky brightness observed through the 
broad, tilted filters (triangles with error bars) and through the 
circular-variable filter (filled circles). Each data point presents the
observed sky brightness after all corrections described in the text,
divided by the COBE/DIRBE zodiacal light model appropriate for its
wavelength and observing date and direction. 
\label{filter_data}}
\raggedright

\subsubsection{Spectra}

The stray-light-corrected, transient-corrected, dark-subtracted data cubes 
were averaged at each wavelength step of the CVF. The pixel-to-pixel 
vignetting variation was normalized by dividing each image by a carefully-reduced
zodiacal light observation obtained during a dedicated calibration orbit
during the mission
\citep{bivianoflat}.
The final spectra were generated by averaging over the central 
$12\times 12$ pixels ($2.4^\prime \times 2.4^\prime$).
Figure~\ref{figzodyspec} shows the results. The spectra are generally
similar to each other, modulo a scale factor. They all show a
similar structure in the 9--11 $\mu$m range.

\plotone{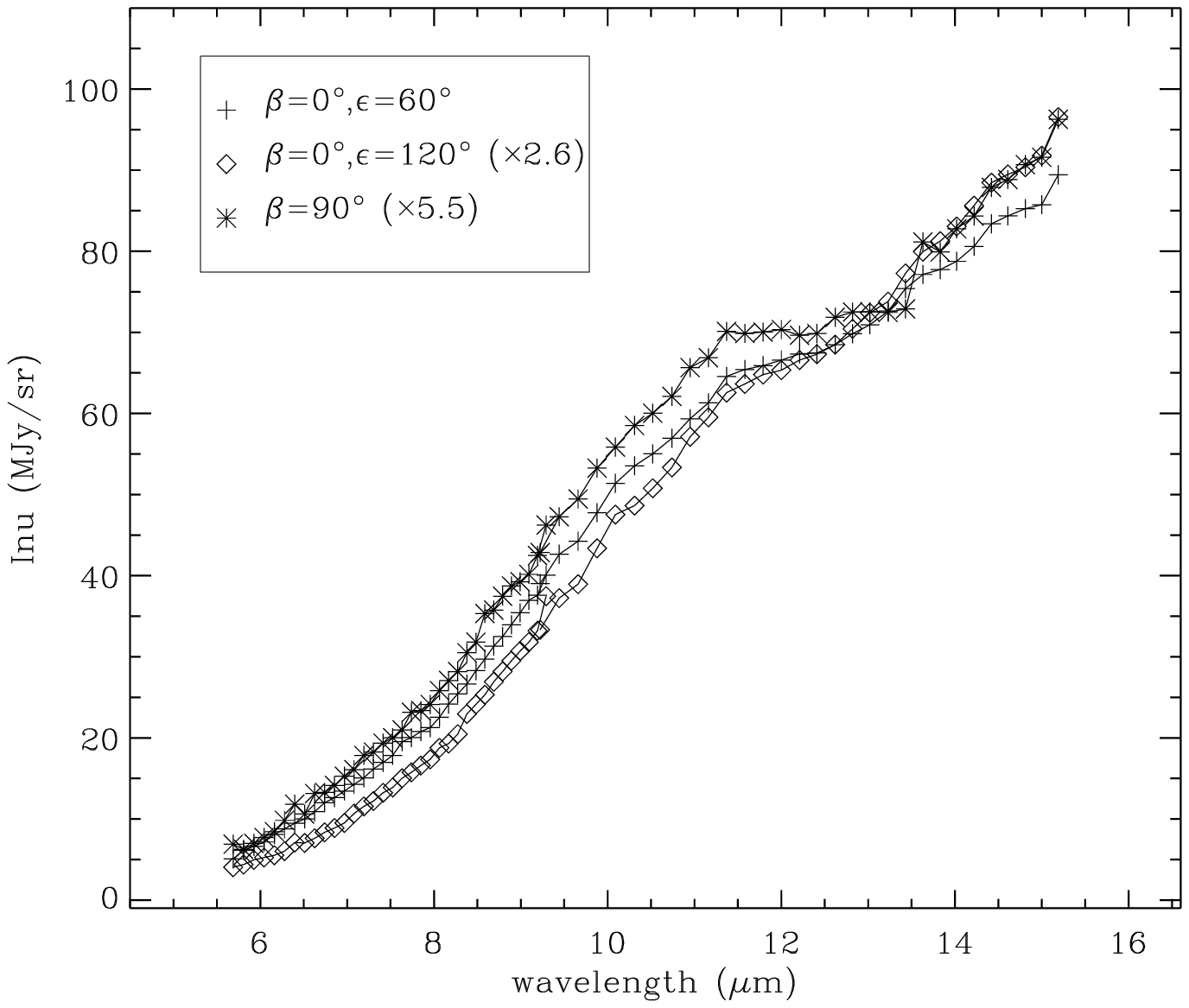}
\figcaption{Spectra of the zodiacal light taken on three lines of sight
representing the extremes of the {\it ISO} operational pointing zone:
in the ecliptic plane at 60$^\circ$ and 120$^\circ$ solar elongation,
and toward the north ecliptic pole. The spectrum in the ecliptic
toward $\epsilon=120^\circ$ was scaled by a factor of 2.6, and the
spectrum of the ecliptic pole was scaled by a factor of 5.5 in 
order to fit on the same scale as the $\epsilon=60^\circ$ spectrum.
\label{figzodyspec}}
\raggedright

\subsubsection{Reduction of archival data}

The reduction of the archival spectra used the final, revolution- and 
position-in-orbit-dependent dark current calibration for the 6$''$ lens
\cite{blommcvf}, as well as the final spectral calibration and
stray light correction described above. 
The fields were mostly clean of interstellar cirrus, and they spanned a
useful range of ecliptic coordinates and solar elongation.
Three of the lines of sight (numbers 16, 15, and 12
in Table~\ref{isolog}) pass through 
significant amounts of interstellar medium.
One of them (number 16) is at a relatively low galactic latitude of 
13$^\circ$, so interstellar contamination is not surprising. 
The other lines of sight are at high galactic latitude; however,
two of the lines of sight
(numbers 15 and 12 in Table~\ref{isolog}) pass through high-latitude molecular clouds 
evident in far-infrared excess \citep{reach98}. In the analysis 
below, we will separate the three lines of sight contaminated by the
interstellar medium (ISM) form the clean, zodiacal spectra.
The spectra are shown in ~\ref{figlutzspec}.

\plotone{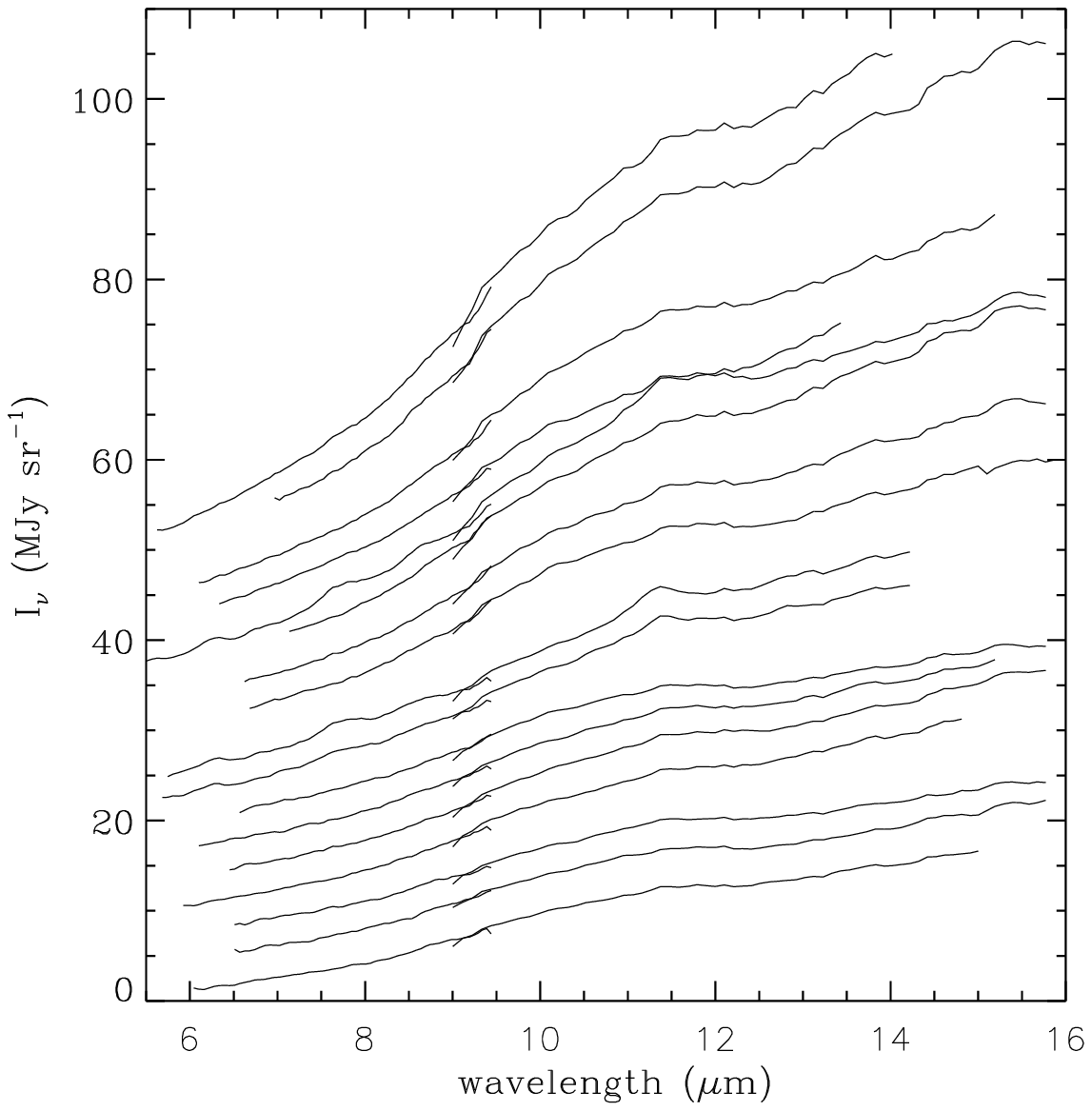}
\figcaption{Spectra obtained from the deep archival observations with the
$6''$ pixel field of view lens. Each spectrum is toward a different
solar elongation and latitude. They were sorted in order of brightness
and offset vertically from each other by 3 MJy sr$^{-1}$ for clarity.
\label{figlutzspec}}
\raggedright

\subsection{Exozodiacal light}

\subsubsection{CAM Observations}

We reanalyzed the archival observations of $\beta$ Pic that were
first presented by \citet{lagage}. The data were taken in a
manner similar to
the zodiacal light observations described above, except the highest
magnification ($1.5''$ pixel field of view) lens was used. The analysis
of the data cube (images at each wavelength) was also similar. To
derive the spectrum of $\beta$ Pic, we measured the flux within a circular 
aperture ($10.5''$ diameter) after subtracting an annular average background
(to $22.5''$ diameter) at each wavelength.

\subsubsection{SWS Observations}

Observations of $\beta$~Pic have been obtained with the ISO SWS
\citep{degraauw} in AOT6 mode in programs of CWAELKEN (CRYSTAL
and CRYSTAL2) and DBOCKELE (VEGA\_ICE), variously covering the full
2.4$-$45.2 $\mu$m range available between the two grating sections.  
\citet{pantinwaelkens} and \citet{malfait} have presented 
the observations from the CWAELKEN programs.  We have reprocessed the 
three sets of observations with the latest pipeline software (compatible with
the ISO pipeline OLP10.1) and interactive techniques to detect and correct
the effects of cosmic radiation on the scan data and dark currents. These
effects are manifest as spurious jumps in signal levels, and drastic
degradation in S/N ratios over SWS bands 2 (7--12 $\mu$m), 3 (12--39.5 $\mu$m),
and 4 (39.5--45.2 $\mu$m) where the $\beta$~Pic flux levels are already
in the dark-current dominated signal regime.  We have interactively treated these
effects on a detector and scan basis, significantly improving the S/N
ratios and removing artifacts. (Some such artifacts were visible
in the earlier reductions, including
band 4, where \citet{pantinwaelkens} and \citet{malfait} 
substituted SWS Fabry-Perot
data provided serendipitously but with unverifiable calibrations.)
Then we 
coadded the corrected products and reduced these to final 1-D spectra rebinned to 
the resolution of the instrument.
This paper is focused on 
the ISOCAM CVF wavelength range.
We present the 5--16 $\mu$m range of the SWS 
spectrum in Figure~\ref{betapic}.  
We mention in passing the presence of CO and H$_2$O gas
absorption lines near 4.0 and 6.2 $\mu$m, respectively, with strengths 
and amplitudes comparable
to those identified in 51 Oph by \citet{vandenancker}.  

\subsubsection{Spectrum $\beta$ Pic}

Figure~\ref{betapic} shows the spectrum of $\beta$ Pic from 5 to 16 $\mu$m.
The broad-band photometric points obtained with ISOPHOT 
\citep{heinrichsen} and {\it IRAS} are overlaid; they show that
the absolute calibration of the SWS and PHOT spectra agrees.
The IRTF observations of \citet{knacke} and the ISOCAM observations
are clearly different from the SWS observations, with the former
being significantly fainter at wavelengths beyond 10 $\mu$m. 
The \citet{knacke} observations cover enough of the short-wavelength,
pure photosphere part of the spectrum (beyond the left-hand-edge of
Fig.~\ref{betapic}) that we can see that the SWS and IRTF data agree
well for the photosphere. The CAM data
also agree in calibration with SWS at the shortest wavelength 5.92 $\mu$m.
The differences between the spectra, which begin around 8 $\mu$m and
become very large at longer wavelengths, are due to the circumstellar
material.
The dust disk begins to dominate over the 
stellar photosphere beyond 8 $\mu$m wavelength; by 10 $\mu$m, the 
circumstellar material is as bright as the photosphere. 
The aperture of the SWS and PHOT observations is significantly larger
than that of the CAM and IRTF observations, and the colder part of the
debris disk is clearly extended. The disk is clearly resolved in
ground-based 10 $\mu$m observations \citep{pantin}, but its extent
is only a few arcsec. The warm, inner disk is completely contained
within all of the apertures discussed here. But at longer wavelengths,
the disk is clearly extended, with a full width at half-maximum 
of $8.7\times 2.9''$ at 25 $\mu$m \citep{heinrichsen}. 
The disk was also resolved in the longer-wavelength CAM CVF images
\citep{lagage}. Thus the CAM and IRTF observations detect primarily
the inner disk, while the SWS and PHOT observations contain essentially
the entire disk including the colder emission.

\epsscale{0.85}
\plotone{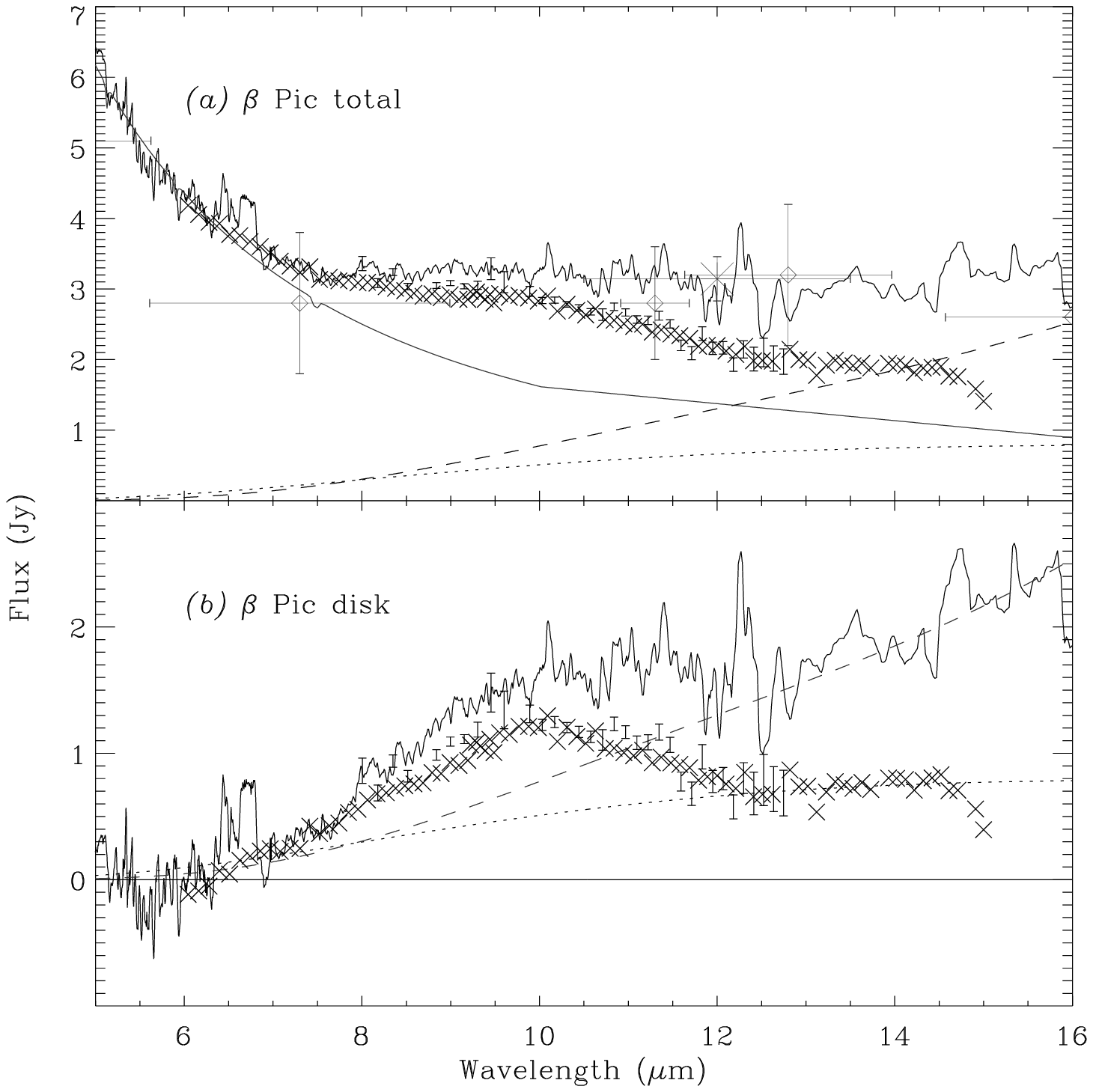}
\epsscale{1}
\figcaption{
{\it (a)} Spectrum of $\beta$ Pic obtained with the ISO SWS (jagged curve),
CAM ($X$ symbols) and IRTF (I symbols).
The broadband photometry from ISO PHOT (diamonds with large error bars) and 
IRAS (asterisks with error bars) are overplotted in grey. 
The Kurucz model stellar photosphere, normalized to the 2.4--4 $\mu$m 
ISO SWS flux, is shown as a solid line.
Approximate models for the debris disk spectra are shown for the
SWS spectrum (dashed line) and the CAM and IRTF spectra (dotted line).
The differences between the observed spectra are primarily due to different 
observing apertures, as the SWS and PHOT data were taken with a large
aperture that includes the entire cold debris disk.
{\it (b)} Circumstellar spectra of the $\beta$ Pic disk obtained by 
subtracting the stellar photosphere from the observed spectra. 
The curves and symbols are the same as in panel {\it (a)}.
The double-peaks at 6.4--6.8 $\mu$m in the SWS spectrum are due to H$_2$O.
\label{betapic}}
\raggedright

The lower panel in Figure~\ref{betapic} shows the
disk spectrum, obtained by subtracting the Kurucz 1993 ATLAS model,
kp00\_8250(g45) scaled to match the flux at 2.5--3 $\mu$m,
for the stellar 
photosphere. The disk spectrum is complicated, being due to a combination
of particles both far from the star, which produces the extremely bright
far-infrared emission peaking around 60 $\mu$m \citep{heinrichsen},
and warmer dust that produces the mid-infrared emission \citep{pantin}.
We will discuss these spectra and their decompositions below in \S~\ref{exosec}.

\section{Results: Continuum plus silicate feature in the zodiacal light}

The mid-infrared zodiacal spectra are dominated by a smoothly increasing
brightness as a function of wavelength, due to thermal emission from
dust at $\sim 270$ K, with a broad shoulder at 9--11 $\mu$m. 
To remove the continuum from the spectra, we used a model for the
zodiacal light that predicts the surface brightness as a function of
wavelength for each target position and observing date (\S\ref{secdark}).
We also fitted each spectrum to a single blackbody in order to
characterize the continuum independent of the model. Table~\ref{conttab}
shows the temperature and optical depths of the continuum spectral fits.

\begin{table}[th]
\caption{Blackbody fits and color ratios}\label{conttab} 
\begin{center}
\begin{tabular}{lcccc} \hline\hline
Spectrum & $T$ (K) & $10^7 \tau$ & $\Delta_{sil}$ \\
\cline{1-4}
ZODY\_60   & $268.5\pm 0.4$  & $2.45 \pm 0.02$ & $5.5\pm 0.1$ \\
ZODY\_120  & $244.1\pm 0.6$  & $1.48 \pm 0.02$ & $5.2\pm 0.7$ \\
ZODY\_NEP  & $274.0\pm 1.1$  & $0.44 \pm 0.08$ & $7.0\pm 2.0$ \\
LUTZ$^b$   & $269.6\pm 5.6$  & $0.92 \pm 0.40$ & $6.9\pm 1.0$ \\
LUTZ$^c$   & $267.1\pm 0.4$  & $0.93 \pm 0.01$ & $8.2\pm 0.7$ \\
\hline
\end{tabular}
\end{center}
\noindent $^b$ average $\pm$ rms from the deep Lutz spectra (excluding
those contaminated by interstellar medium)

\noindent $^c$ weighted average and statistical uncertainty from deep Lutz spectra 
(excluding those contaminated by interstellar medium)
\end{table}

\figcaption{Comparison of the ISOCAM spectra (symbols, as in
Fig.~\ref{figzodyspec}) with the DIRBE zodiacal light model for
the same coordinates and date (large errors bars and solid line).
\label{figzodydirbe}}
\plotone{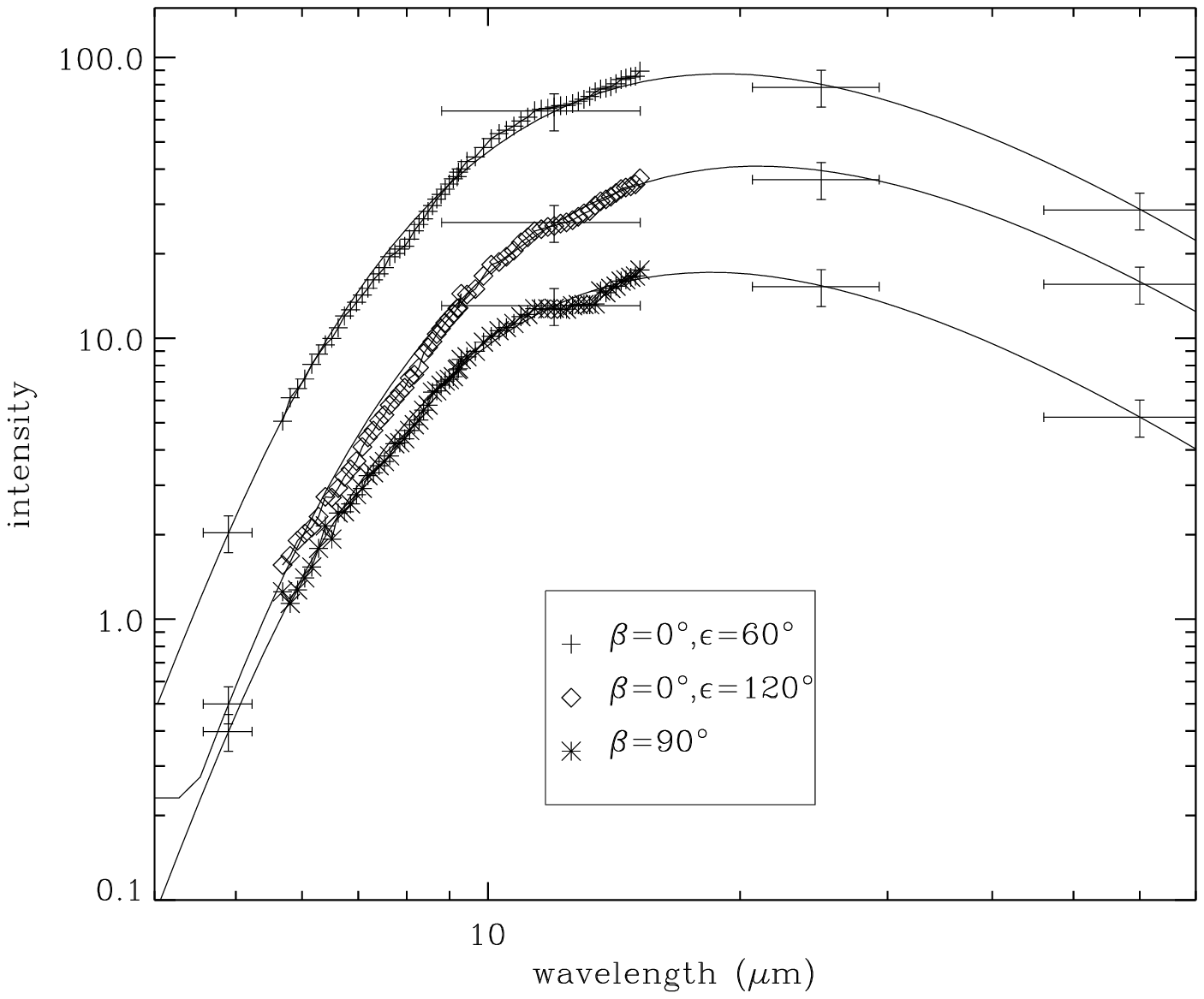}
\raggedright

\figcaption{Ratio of the observed zodiacal light spectra from
Fig.~\ref{figzodyspec} to a blackbody continuum. The continuum
was fit using only wavelengths outside of 9--11 $\mu$m. 
\label{figzodyratio}}
\plotone{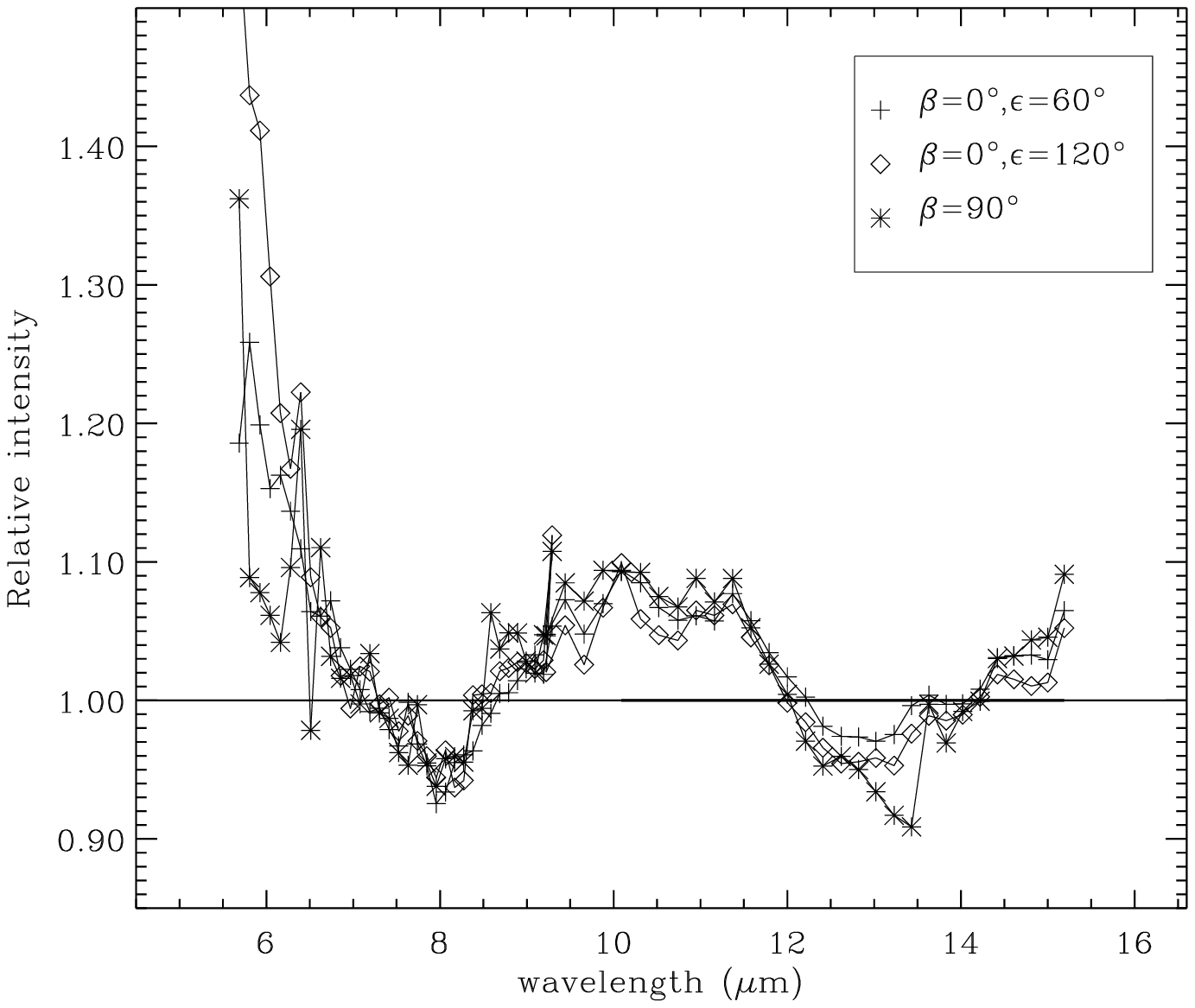}
\raggedright

Figure~\ref{figzodydirbe} compares the observed spectra to the
their corresponding DIRBE zodiacal light model predictions. 
Figure~\ref{figzodyratio} shows the observed spectra divided by 
their corresponding DIRBE zodiacal light model predictions. 
Because the model includes only a blackbody emissivity, the ratio
spectrum should reveal all structures in the spectrum---both narrow
lines and broad bands. It is clear that the model
very accurate predicts the spectra: it matches to within 10\%, with no
adjusting of the model to fit the new ISOCAM data. 
Because of the excellent agreement between the
COBE/DIRBE model and the ISOCAM data, and because the
COBE/DIRBE data span a much wider parameter space of spatial 
and wavelength coverage, we need not delve into a detailed
model for the zodiacal brightness distribution in this
paper, but instead we refer the reader to \citet{kelsall}.
The value of the new results presented here is in measuring the
spectrum of the zodiacal light at a spectral resolution sufficient
to detect spectral features.

The ratio of the observed (ISOCAM) to model spectra
shows clear systematic structures. This structure consists of a 
well-defined, broad peak at 9--11 $\mu$m, 
superposed on an overall `bowl'-shaped residual. 
The 9--11 $\mu$m feature is the same structure as was seen in our
earlier mid-infrared zodiacal light spectrum \citep{reach96}, but
now we have sufficient signal-to-noise and confidence in 
the instrumental calibration to confirm a detection. 
The 9-11 $\mu$m excess has also been detected in data from the 
Mid-Infra re Spectrometer (MIRS) on the Infrared Telescope in Space
(IRTS) by \citet{ootsubo98}, although the MIRS data extend in
wavelength only to 11.59 $\mu$m, which was not quite long enough to
see the return from the feature to the continuum at long wavelengths.
The `bowl' shape
is due to the DIRBE model not accurately predicting the shape of
the continuum. As discussed below, the zodiacal light emissivity 
is a superposition of emission from particles of many sizes (each
with their own temperatures) as well as particles at many locations
(also having their own temperatures). The DIRBE model incorporates
the spatial dependence of the temperature, but it assumes a single
blackbody at each location, which is not strictly true. Because
the overall shape of the continuum is best studied with data covering
an even wider range of wavelengths than ISOCAM, we will emphasize
the 9--11 $\mu$m feature in this paper and briefly discuss the
continuum shape in comparing with the wider DIRBE wavelength coverage.

\epsscale{0.8}
\plotone{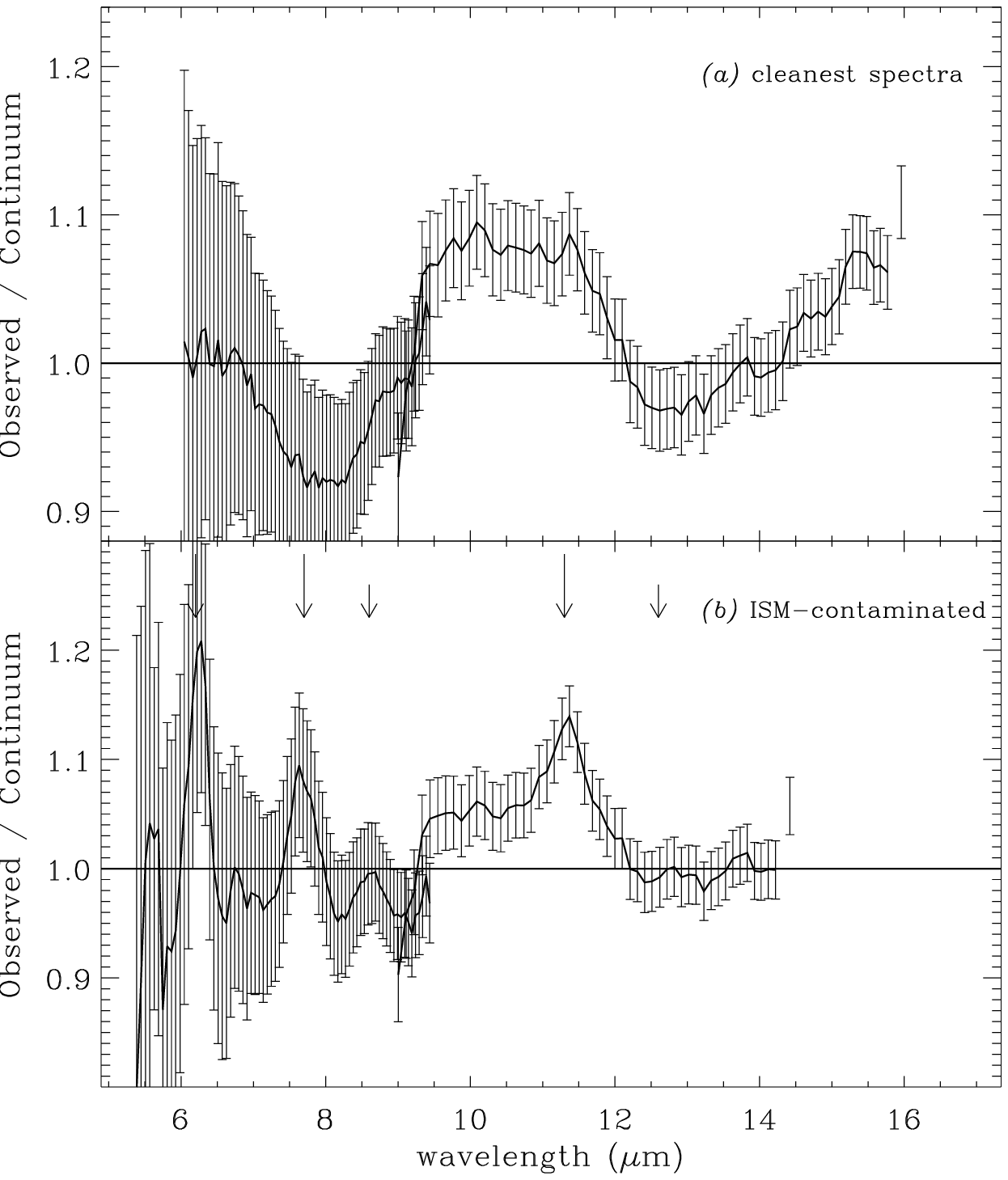}
\epsscale{1}
\figcaption{Ratio of the archival zodiacal light spectra, taken with
the $6''$ lens and CVF of ISOCAM, to a zodiacal light model.
The top panel shows the average of the observed to continuum ratios
for the 14 spectra with no evidence for interstellar contamination.
The bottom panel shows the average for the 3 spectra with 
interstellar contamination.
The ISM lines, attributed to polycyclic aromatic hydrocarbons
(PAH), are labeled with vertical arrows.
\label{figlutzrat}}
\raggedright

Figure~\ref{figlutzrat} shows the average of the archival spectra divided 
by their continuum from the DIRBE zodiacal light model. 
The three ISM-contaminated spectra were treated separately. They
clearly show bright infrared bands at 6.2, 7.7, and 11.3 $\mu$m. 
These are the brightest emission bands in spectra of diffuse 
interstellar matter, and they are widely attributed to 
polycyclic aromatic hydrocarbons (PAH) \citep{boulrhooph}. 
The uncontaminated spectra show no such interstellar
lines, but instead they are dominated by the same broad band at 9--11 $\mu$m
that was seen in the dedicated zodiacal light observations. One
straightforward result of the ISOCAM observations is that there are
no PAH bands from the zodiacal light. The PAH bands are bright from
the interstellar medium because of the presence of very small grains
and a pervasive UV-bright radiation field that transiently heats
the grains to high temperatures. In the Solar System, the interval
between successive photons is shorter, and the particles will be close
to their equilibrium temperature. This temperature is still hot enough
so that the mid-infrared bands would be excited. The upper limit to
PAH bands from the zodiacal light is $< 2$\% of the continuum 
brightness.

All of the zodiacal spectra, including the dedicated observations
toward the ecliptic plane and pole (Fig.~\ref{figzodyratio}), the
deep, archival observations (Fig.~\ref{figlutzrat}) and 
observations through the wide-band filters (Fig.~\ref{filter_data}),
show an excess in the 9--11 $\mu$m range
that is characteristic of silicates. 
To characterize the brightness of the silicate feature, we define the
quantity 
\[
\Delta_{sil} = 100 \left(
   \frac{ \langle I_\nu  \rangle_{(9-11 \mu{\rm m}}\rangle }
        { \langle \tau B_\nu (T) \rangle_{(9-11 \mu{\rm m}}\rangle }
	  - 1\right), \]
which is the percentage excess of the observed spectrum $I_\nu$ relative to the
blackbody fit $\tau B_\nu(T)$ (where $\tau$ and $T$ are the fitted optical
depth and temperature and $B_\nu$ is the Planck function) 
averaged over the wavelengths of the feature.
Table~\ref{conttab} shows the values of $\Delta_{sil}$ for the zodiacal spectra.
To first order, the continuum and 9--11 $\mu$m feature shapes and amplitude are 
comparable for all of the spectra. 
The amplitude of the 9--11 $\mu$m excess seen in the IRTS/MIRS data 
\citep{ootsubothesis} is similar to that of our data.
We present a more detailed comparison 
of the spectral shapes and systematic dependence on the viewing geometry in
section~\ref{sec_fit}.

\def\extra{
\section{Comparison to other experiments}

COMPARE TO DIRBE OVER WIDER WAVELENGTH RANGE

COMPARE TO JAPANESE 

COMPARE TO GERMANS

\placefigure{figzodydirbe}
}

\section{Comparison to theoretical models}

\subsection{Compositions}

We calculated the infrared emission spectra for a range of 
particle compositions and size
distributions, to learn what kinds of particles can create
the observed zodiacal light spectra.
Table~\ref{comptab} describes the particle compositions and optical 
constant data that we used.
The calculations are an update of those that
were performed to interpret the 12--100 $\mu$m broad-band colors of
the zodiacal light \citep{reach88}. 
We are restricting our calculations to plausible materials.
The earlier calculations showed that pure, crystalline
silicates are too cold (because they are nearly transparent to
sunlight) while carbonaceous dust is too hot (because it absorbs
sunlight so well). Furthermore, it is unlikely that extremely
pure solids will exist in comets. Solids in the Solar System formed from 
interstellar dust, so we consider interstellar silicate dust as an
indicator of the purity of the building blocks. 
A model for the optical constants of `astronomical silicates'
\citep{dl84,laor} was found to be reasonably approximated by
a volume mixture of pure andesite and 3\% graphite \citep{reach88}.
Extensive infrared observations of C/Hale-Bopp demonstrated that
even the silicate dust from a new comet is impure: pure silicate
temperatures were too low 
to produce the observed continuum shape and silicate feature 
ratios \citep{grun01}. (We confirm that this is true for many pure
silicates; however, some amorphous silicates at 2.82 AU from the Sun
have temperatures that are in the range [$T\sim 180$ K] required 
to explain the Hale-Bopp spectrum without need for extra absorption.)
For this paper, we `dirtied' the pure silicates to the same
degree as the `astronomical silicate' model of \citet{dl84}.
Specifically, we use the Maxwell-Garnet mixing rule to add
amorphous carbon with a volume filling factor $f_{car}$ 
(listed in Table~\ref{comptab}) such
that the Fresnel transmittance averaged over the wavelength 
range 0.2--0.7 $\mu$m (where sunlight does most of its heating)
is at least $T_{vis}\geq 0.93$, the value for astronomical silicate.

\begin{table}[th]
\caption{Particle compositions and optical data}\label{comptab} 
\begin{center}
\begin{tabular}{lccl} \hline\hline
Name & wavelength range & reference & $f_{car}^a$\\
\cline{1-4}
astronomical silicate {\it (a)}
	& 0.001--1000 $\mu$m & \citet{laor} & ... \\ 
amorphous olivine (Jena)
	& 0.2--500 $\mu$m & \citet{dorschner95} & 0 \\ 
amorphous forsterite
	& 0.043--61.7 $\mu$m & \citet{scott96} & 0.1\\
amorphous pyroxene 
	& 0.2--500 $\mu$m & \citet{dorschner95} & 0\\ 
crystalline olivine {\it (c)}
	& 0.01--2980 $\mu$m & \citet{mukai90} & 0.05\\ 
crystalline pyroxene {\it (d)}
	& 6.66-487 $\mu$m & \citet{jaeger98} & 0.033\\ 
crystalline enstatite $\parallel$
	& 0.043--99 $\mu$m & \citet{jaeger98} & 0.58\\ 
crystalline enstatite $\perp$
	& 0.043--99 $\mu$m & \citet{jaeger98} & 0.58\\ 
montmorillonite {\it (e)}
	& 0.023--2000 $\mu$m & \citet{montmorillonite} & 0.46\\ 
glassy carbon 
	& 0.2--100 $\mu$m & \citet{edoh} &  ... \\
amorphous carbon {\it (f)}
	& 0.105--800 $\mu$m & \citet{preibisch} & ...\\ 
SiC
	& 0.001--1000 $\mu$m & \citet{laor} & ... \\ 
Mg-rich crystalline olivine 
	& 0.2--8000 $\mu$m & \citet{fabian} & 0.37 \\ 
\hline
\end{tabular}
\end{center}
\end{table}

\subsection{Absorption efficiencies}

For each particle composition, the absorption efficiency was computed on the 
same wavelength grid on which the laboratory optical measurements were provided. 
The computations were performed for particles radii on a logarithmic
grid from 0.11 to 1000 $\mu$m (particle mass $10^{-14}$ to $10^{-2}$ g).
The calculations used Mie theory, and they were somewhat computationally
intensive due to the inclusion of very large particles.
The grain temperatures we calculated
by balancing the solar radiation at 1 AU by the integrated thermal emission
of the grains. The grains are assumed to be isothermal, rapidly rotating
spheres; thus, absorption was assumed to occur into an effective
cross-section $\pi a^2$, and emission out of $4\pi a^2$, where
$a$ is the particle radius. The largest grains can conceivably support
a small temperature gradient, meaning they would emit at slightly higher
temperatures than we have calculated. 


\subsection{Size distributions}

The volume emissivity is calculated by integrating over the size distribution
\begin{equation}
{\cal E_\nu} = \int{ \frac{dn}{da} B_\nu[T(a)] \pi a^2 Q_{abs}(a) da},
\label{emiss_equation}
\end{equation}
where $dn/da$ is the size distribution, $B_\nu$ is the Planck
function, $a$ is the particle radius, and $Q_{abs}$ is the absorption
efficiency. In practical units ${\cal E_\nu}$ is in MJy~sr$^{-1}$~AU$^{-1}$
or in cgs units, erg~s$^{-1}$~Hz$^{-1}$~cm$^{-3}$~sr$^{-1}$.
For our calculations, we use the following size distributions.
The `interplanetary' and `lunar' size distributions are
based on a combined analysis of particle detectors on Earth-orbiting 
satellites, meteors, and lunar microcraters; the only difference between
these distributions is for the smallest particles, where the `lunar'
distribution uses the raw lunar microcrater data, while the `interplanetary'
distribution uses an estimate based on $\beta$-meteoroid 
production \citep{grun85}. The `interplanetary' size distribution
remains the best estimate of the size distribution of cosmic
material at 1 AU from the Sun, and it has recently been shown
to be consistent with the sizes of particles inferred from 
the {\it Long Duration Exposure Facility} ({\it LDEF}) spaceward
face \citep{mcdonnell98}. Three other size distributions are included
for comparison. The `interstellar' size distribution is a power-law,
$dn/da\propto a^{-3.5}$,
as inferred from the interstellar extinction curve \citep{mrn,dl84};
this is not intended to represent the size distribution of actual
interstellar grains in the Solar System---the Ulysses and Galileo
experiments have clearly shown that small interstellar grains do
not penetrate into the inner Solar System \citep{landgraf}---but
rather it is included as a simple model for comparison.
The `halley' size distribution is based on analysis of the particle
detector data from the P/Halley encounter \citep{lamy_halley}; it
may represent the input size distribution for cometary particles
that contribute to the zodiacal dust cloud. Similarly, the
`hanner' size distribution, based on ground-based coma observations of
short-period comets \citep{hanner84}, represents freshly-produced
cometary material.

To determine which particles actually produce the zodiacal emission, we
show the integrand from equation~\ref{emiss_equation}, in 
Figure~\ref{emiss_size}. 
At wavelengths of strong material resonances, small particles 
contribute, but larger particles (10--100 $\mu$m radius) dominate 
the continuum, especially at long wavelengths.
The size distribution is therefore
a significant determinant of the strength of spectral features.
For size distributions with relatively more small particles,
the short-wavelength emission, as well as the spectral feature emission,
can have a significant small-particle
contribution, because the small particles are (for at least
moderately-absorbing materials) hotter than the larger particles.

\plotone{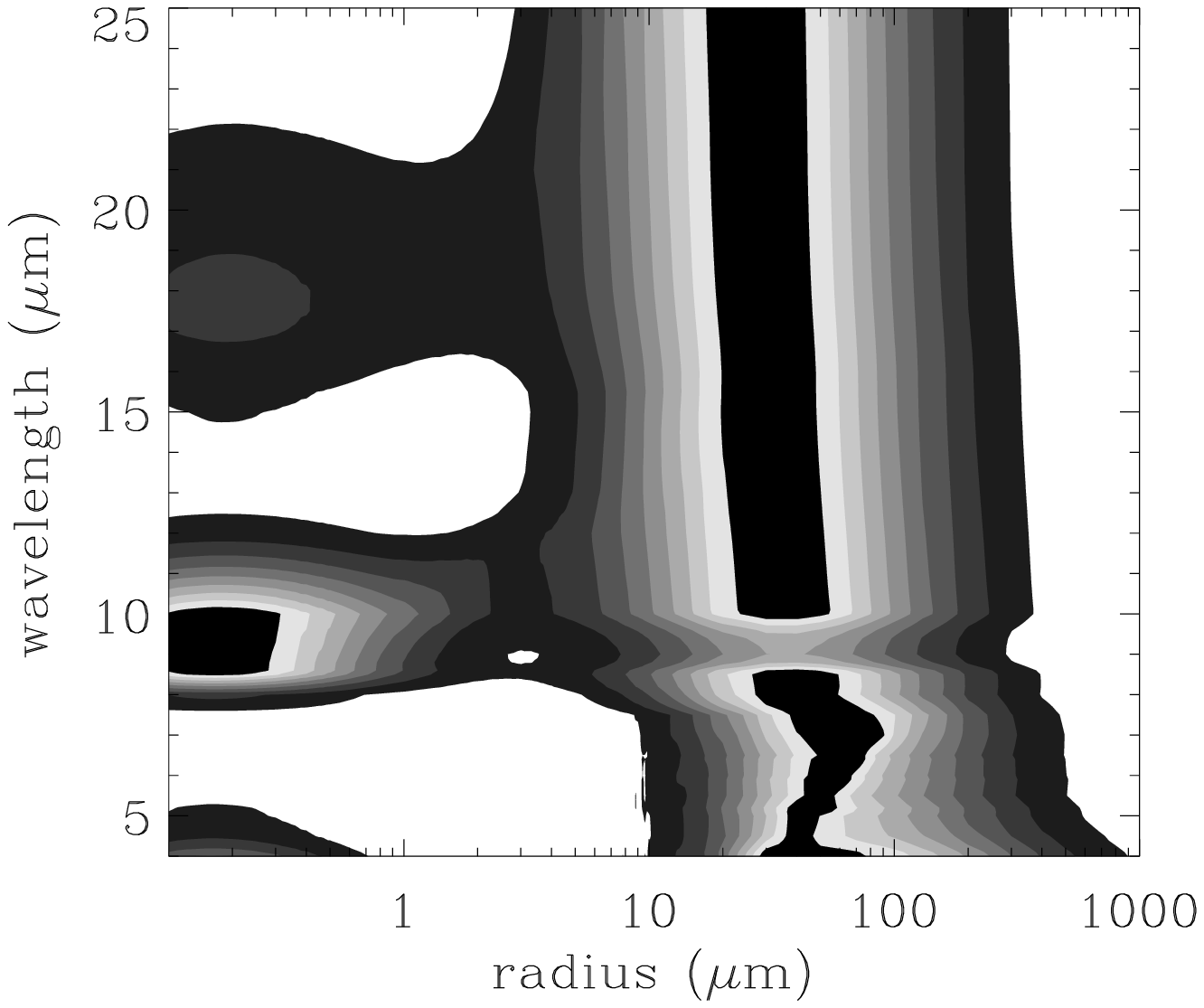}
\figcaption{The fractional contribution of pyroxene grains of various
sizes to the zodiacal light at a range of wavelengths. 
These data are for the amorphous pyroxene model and the interplanetary size
distribution. For each wavelength, the emissivity integrand as
a function of particle size (specifically, 
$(dn/da) B_\nu[T(a)] \pi a^2 Q_{abs}(a) a$ was normalized 
to its peak value. At most wavelengths, particles in the size range
10--100 $\mu$m produce the bulk of the emission. But at the wavelengths
of the silicate feature, a significant proportion of the emission 
comes from very small particles that are not opaque at 10 $\mu$m.
The rather complicated dependence of the emissivity on the particle 
size and wavelength makes it impossible to use single-temperature 
or single-size models to fit the entire spectrum.
\label{emiss_size}}
\raggedright

\subsection{Comparison of models to observations}

Model predictions for the zodiacal spectrum are shown in Figure~\ref{plot_emiss}.
The models can be characterized to first order as blackbody continua plus
(in many cases) silicate features at 9--11 $\mu$m. Using the same fitting 
procedure that was used on the data, we measured the color temperature
of the continuum over the ISOCAM wavelength range; Table~\ref{modtemptab}
shows the results. Please note that these color temperatures are only 
valid over the 5--16 $\mu$m wavelength range: extrapolating these fits to
longer wavelengths ({\it e.g.} $\lambda>20$ $\mu$m) results in very poor 
predictions of the spectrum.
The carbonaceous materials are generally too hot to be the
dominant, individual material producing the zodiacal light. The crystalline
silicates were generally too cold, but after they were
`dirtied' by adding carbonaceous material (as described above) some of their
temperatures are comparable to the observed zodiacal spectra.
All of the models with the \citet{grun85} interplanetary size distribution have
approximately the observed color temperature, because large, opaque particles
dominate the emission. 

\begin{table}[th]
\caption{Color temperature of model continuum$^a$}\label{modtemptab} 
\begin{center}
\begin{tabular}{lccccccc} \hline\hline
& \multicolumn{7}{c}{Size Distribution}\\
\cline{2-8}
Material              &     hanner &  power &   grun &  lunar & halley-la & halley-mc & halley-ks \\
\cline{1-8}
astronomical silicate     &  242.2 &  250.0 &  264.2 &  264.9 &  237.6 &  247.7 &  239.5 \\
amorphous olivine         &  236.1 &  248.6 &  252.1 &  254.9 &  229.6 &  244.5 &  232.9 \\ 
dirty amorphous enstatite &  351.9 &  420.2 &  275.1 &  318.8 &  322.0 &  399.8 &  315.4 \\ 
crystalline olivine       &  280.6 &  289.5 &  269.3 &  271.9 &  270.6 &  296.1 &  274.1 \\ 
dirty montmorillonite     &  372.9 &  446.2 &  273.8 &  325.4 &  336.2 &  432.3 &  328.2 \\ 
glassy carbon             &  404.8 &  547.4 &  290.5 &  373.8 &  361.4 &  483.4 &  348.3 \\
\hline
\end{tabular}
\end{center}
\noindent $^a$ temperature of best-fitting blackbody for wavelengths 6--16 $\mu$m
excluding 8--13 $\mu$m
\end{table}

\epsscale{0.85}
\plotone{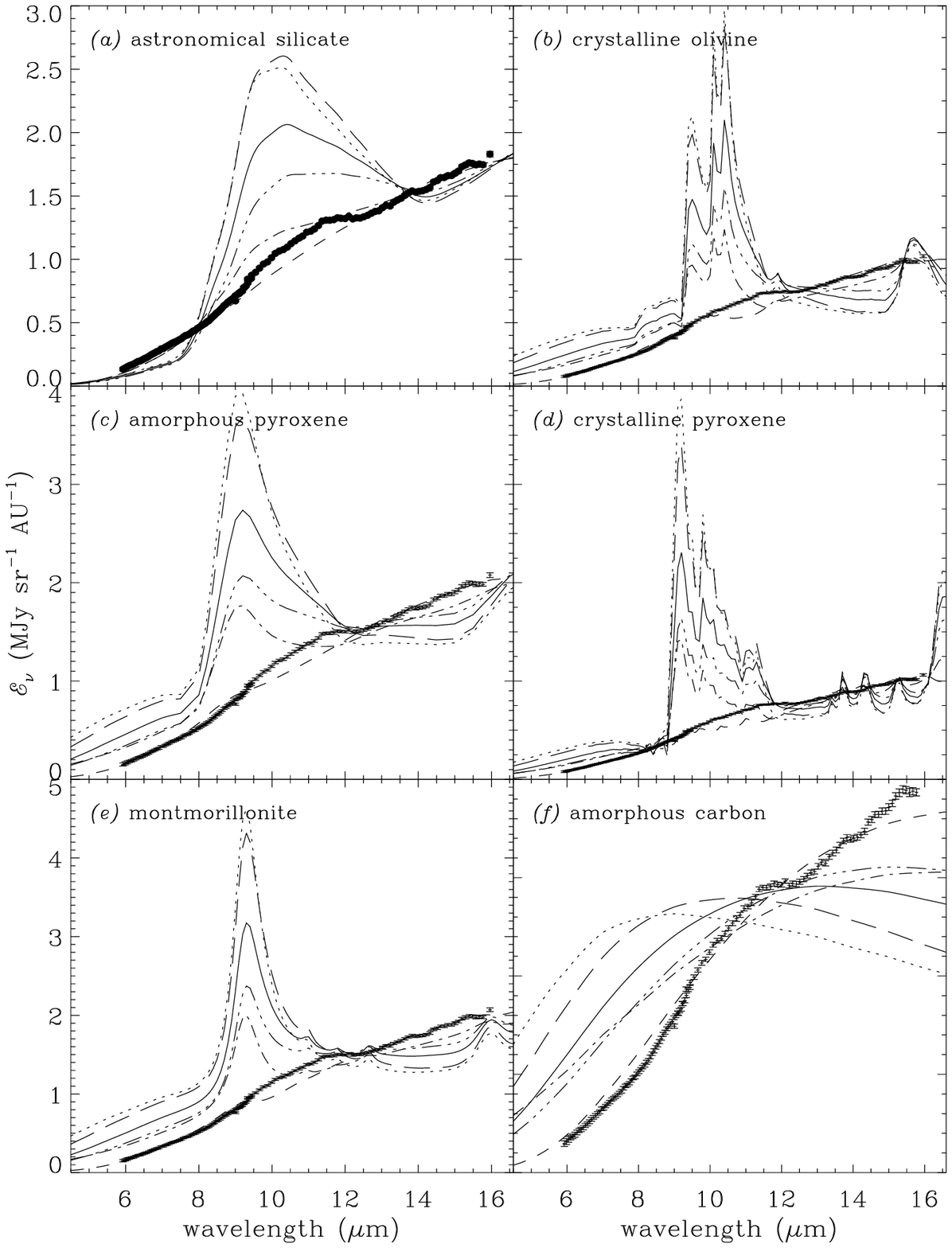}
\epsscale{1}
\figcaption{Model spectra of interplanetary dust particles at 1 AU
from the Sun, for different compositions and size distributions.
Each panel shows the models for one particle composition and 5 different
size distributions 
(long dash: halley/McDonnell,
dotted: interstellar,
solid: Hanner,  
dash-dot-dot-dot: halley/Lamy, 
dash-dot: lunar, 
dashed: interplanetary;
listed in order of silicate feature strength for astronomical silicate).
The `power-law' size distribution, which has the largest abundance of
small particles, has the highest-contrast spectral features, while the
\citet{grun85} interplanetary size distribution, which has the lowest
abundance of small particles, has essentially no silicate feature.
Panel {\it (a)} represents interstellar material;
panels {\it (b), (c)} represent amorphous silicates; and
panel {\it (d)} is a crystalline silicate;
panel {\it (e)} is a meteorite sample;
and panel {\it (f)} is an amorphous carbon sample.
In each panel, the average observed zodiacal spectrum is overplotted as
a set of thick black circles. It is clear that there are no high-contrast
features in the observed spectrum, ruling out essentially all of the
small-particle-cominated size distributions.
\label{plot_emiss}}
\raggedright

The observed zodiacal spectrum has very little structure; specifically, there are
no features above 10\% of the continuum. In order to assess which model spectra show
comparable features, we divided each model spectrum by its blackbody continuum 
fit (see Fig.~\ref{plot_emiss_sub}).
First, it is evident that all of the spectra, both model and observed,
exhibit a rise at short wavelength relative to their blackbody
fits. This rise can be easily explained by a range of dust temperatures
contributing to the emission. The blackbody fit will yield the color
temperature that dominates the 6--15 $\mu$m wavelength range. It
will not include hotter dust, which dominates at shorter wavelengths
because of the steepness of the Wien portion of the Planck curve. 
The short-wavelength rise can be due to small silicates or to 
carbonaceous material. 

\epsscale{0.85}
\plotone{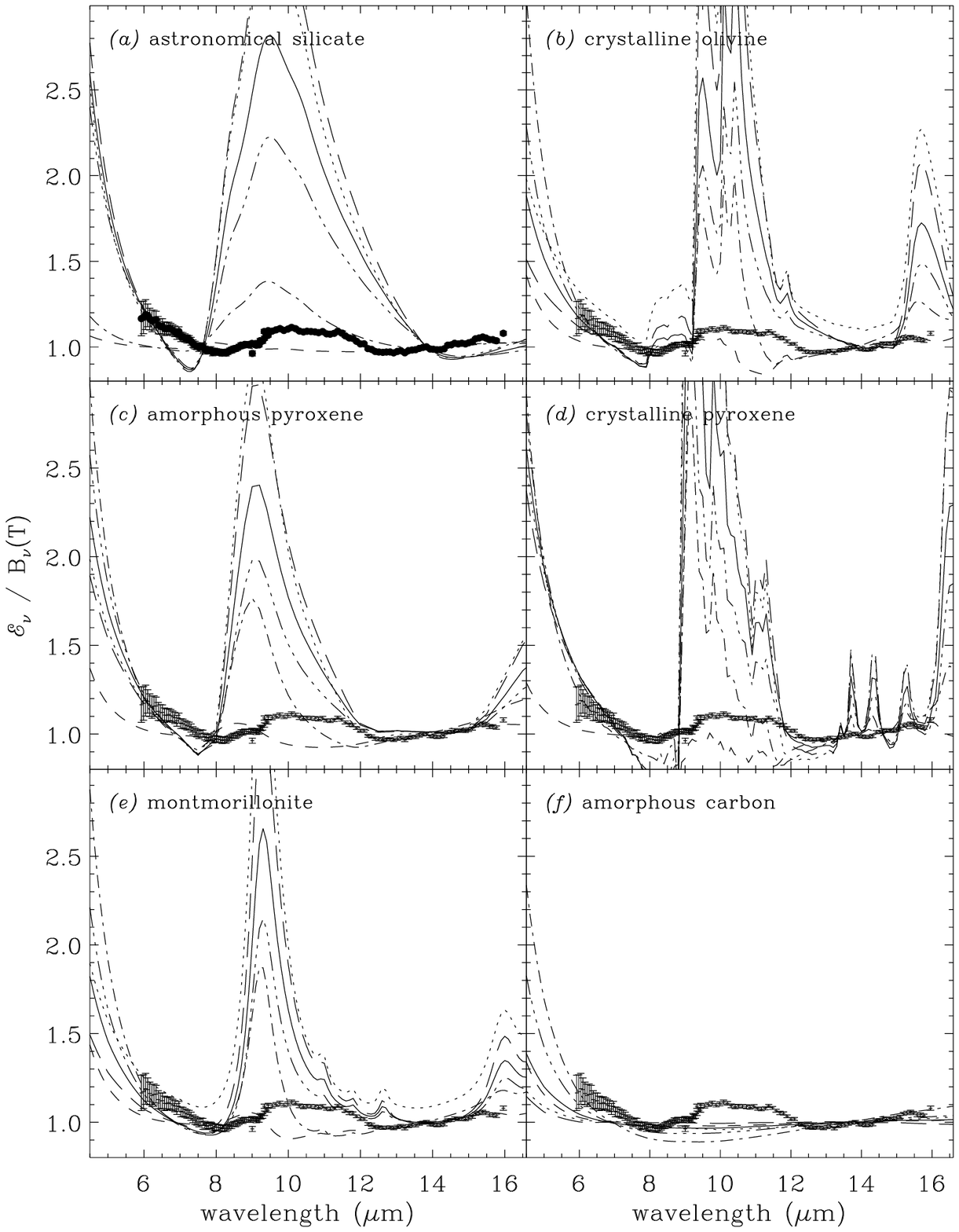}
\epsscale{1}
\figcaption{Model spectra from Fig.~\ref{plot_emiss}, divided by a 
blackbody fit to their mid-infrared continua. The fits were performed
in the same way for the observed data, which are
overplotted on each panel. The rise in the observed ratio spectrum
at short wavelengths is reproduced in the model spectra; it is due to
the spread in temperatures as a function of particle size. The bump
at 9--11 $\mu$m that could be a silicate feature is roughly comparable
to those predicted by the silicate models, but it does not match
any of these models in detail. The zodiacal dust must include a
range of different mineralogies; they cannot be dominated by a single
crystalline or amorphous silicate.
\label{plot_emiss_sub}}
\raggedright

The spectral features predicted by the models are generally much too strong
and too structured to match the observed spectrum, except for the
large-particle-dominated models. In Figure~\ref{plot_emiss_sub}, it
is clear that the individual silicates, both crystalline and amorphous,
generate distinctive shapes in their 9--11 $\mu$m spectral features.
The observed spectral feature is rather broad and smooth, lacking 
sharp peaks. However, the observed line is `boxy' and the short- and 
long-wavelength edges of the observed feature roughly correspond to resonances 
in silicates. We can reach several simple conclusions from these observations.
First, the short-wavelength excess emission seen in the observed spectrum
is more likely due to carbonaceous (or another strongly-absorbing) material;
if it were due to small, hot silicates, then the 9--11 $\mu$m feature would
be much stronger.
Second, the zodiacal light is produced mostly by large ($> 10$ $\mu$m) 
particles, and the
size distributions with large abundances of small particles can
be ruled out. From Table~\ref{disttab}, only the interplanetary and
lunar size distributions from \citet{grun85} are plausible. The other
size distributions, which are based on observations of comets and the
interstellar medium,
all have far too many small particles, making their spectral features
and too prominent (i.e. with $Delta_{sil}$ much larger than observed). 
Thus is is clear that 
zodiacal dust is very different, in size or composition,
from the dust that produces the
observed cometary or interstellar silicate spectral features.

\begin{table}[th]
\caption{Particle size distributions$^a$}\label{disttab} 
\begin{center}
\begin{tabular}{lccl} \hline\hline
Name & $k_{small}$ & $k_{large}$ & reference \\
\cline{1-4}
      interplanetary   & -2.16 & -2.00 & \citet{grun85} \\
     lunar   & -3.74 & -2.05 & \citet{grun85} \\
interstellar & -3.50 & -3.50 & \citet{mrn} \\
    Halley/Lamy   & -2.53 & -3.40 & \citet{lamy_halley} \\
    Halley/McDonnell   & -3.11 & -3.96 & \cite{mcdonnell_halley} \\
    Hanner   & -2.66 & -3.59 & \citet{hanner84} \\
\hline
\end{tabular}

$^a$ $k_{small}$ and $k_{large}$ are the approximate power-law indices
of the size distributions between particle sizes 0.3--2 $\mu$m and
2--20 $\mu$m, respectively.
\end{center}
\end{table}

\subsection{Mixture model}

We constructed a non-unique, but illustrative, 
model that quantitatively matches the broad zodiacal spectrum. 
The model uses a large-particle-dominated size distribution with just enough
small silicates to make the weak 9--11 $\mu$m feature. 
A better fit to the detailed shape of the silicate feature using different 
particle shapes is described below; the model described here addresses
the overall spectral shape including feature and continuum. First, because
the interplanetary size distribution yields no silicate features and
the lunar (and other) size distributions make too strong a silicate feature, 
we had to construct a new size distribution. 
This was simply done by taking the difference between the
model spectra for the interplanetary and lunar distributions, and adding in
a fraction of that difference to the interplanetary distribution until
it makes silicate feature amplitude comparable to the observed one. 
This is valid
because the model is a linear function of size distribution
(eq.~\ref{emiss_equation}), and the lunar size distribution is identical
to the interplanetary size distribution except for small particles
\citep{grun85}. Second, we had to combine different materials in order
to generate a wide silicate feature.

\epsscale{0.7}
\plotone{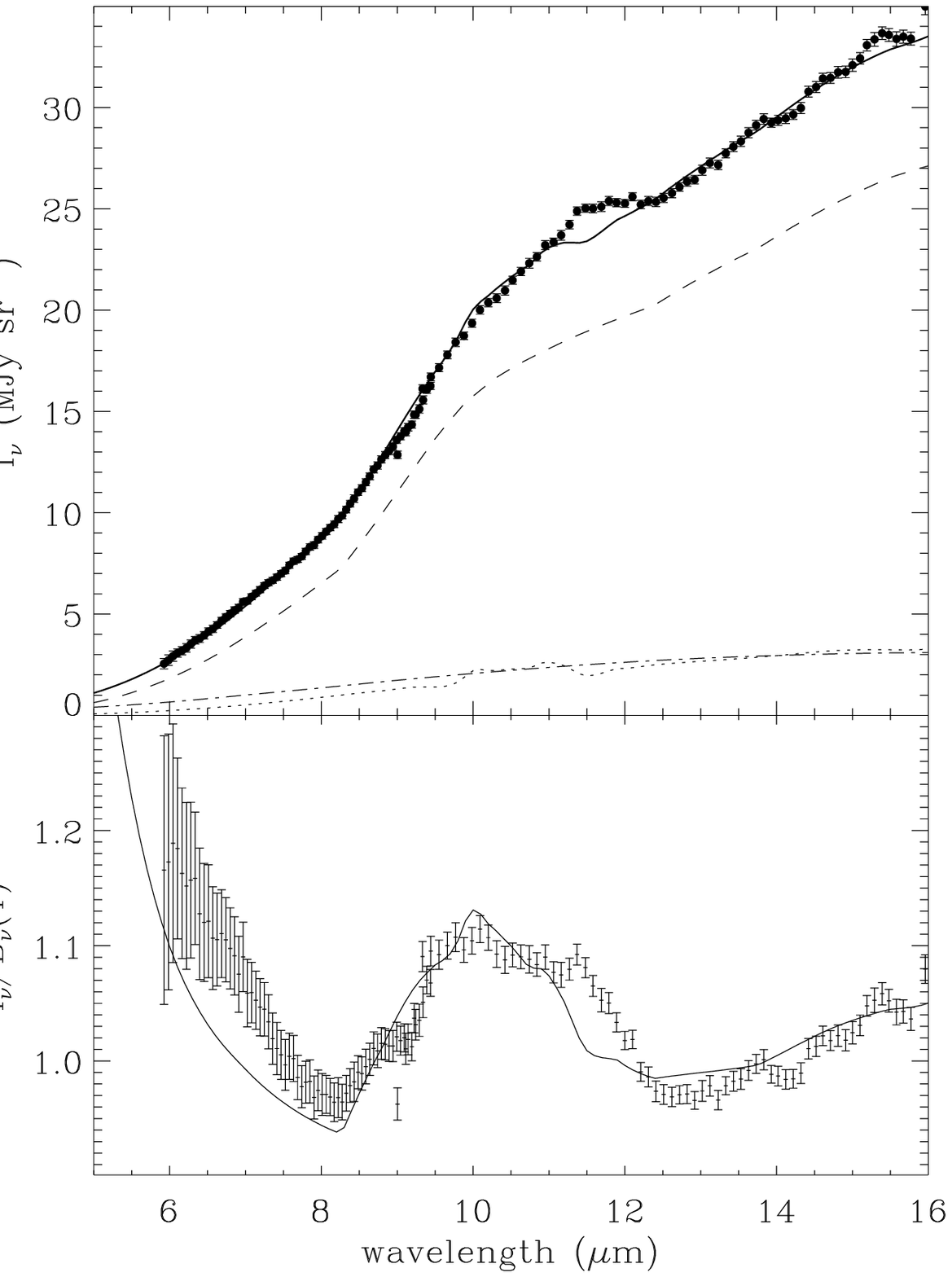}
\epsscale{1}
\figcaption{Comparison of a mix of silicates and carbonaceous materials
to the observed spectrum of the zodiacal light. The top panel shows the
observed spectrum (circles with error bars), together with the mixture
model (solid line) and the brightness of each of the 3 components of
the mixture (dash: dirty amorphous olivine,
dash-dot: glassy carbon,
dash: crystalline olivine).
The bottom panel shows the ratio of the intensity to a blackbody continuum
fit for the observed (symbols with error bars) and mixture model (line)
spectra.
\label{plot_mix}}
\raggedright

Figure~\ref{plot_mix} shows the mixture,
compared to the observed spectrum. 
The fit to the spectral feature is relatively poor,
but the main features of the observed spectrum
are reproduced. The mixture contains three materials: 
amorphous forsterite/olivine from \citet{scott96} [80\% of the 13--15 $\mu$m intensity], 
crystalline olivine from \citet{mukai90} [10\%], and
amorphous carbon from \citet{edoh} [10\%].
Relative to cosmic elemental abundances, this is a silicate and oxygen-rich mixture.
Most of the carbon is mixed in with the amorphous forsterite/olivine \citep{scott96},
which was `dirtied' using a 10\% volume mixture of glassy carbon; the 
free, amorphous carbon grains contain the remaining 35\% of the carbon.
The relative paucity of carbonaceous material could be due to its volatility:
the grains, whether originally cometary or asteroidal, 
have been exposed to sunlight for $> 10^6$ yr \citep{grun85}, which
may be long enough to remove their volatiles.
The elemental abundances from our mixture model should not be taken as the
total elemental abundances of the interplanetary dust, because they are weighted
toward particles that are relatively bright in the mid-infrared. 
The elemental abundances from our mixture model are different
from those of fresh cometary ejecta, which has a significant gaseous component
rich in the lighter elements (e.g. as detected by impact detectors that 
flew through comet Halley's coma in 1986).
The size distribution is dominated by large (10--100 $\mu$m radius) particles.
For crystalline olivine and amorphous carbon, the size distribution
was intermediate (50\% each) between the interplanetary and lunar size distributions,
while the amorphous olivine, which dominates the spectrum, had a size
distribution very close to interplanetary (92\%, with 8\% lunar).

In order to fit the detailed shape of the silicate feature, we
had to consider the effect of particle {\it shape}.
Inspecting Figure~\ref{plot_emiss_sub} carefully, it is evident that none
of the theoretical models for spheres produce silicate features 
extending beyond 11 $\mu$m, as the observed feature does. 
This is true despite the presence of strong resonances in 
crystalline olivine at 11.4 $\mu$m \citep{fabian}.
It is a well-known property of spheres that their absorption cross-sections
do not necessarily peak at the wavelengths of the resonances,  
instead having wavelength dependence, in the small particle approximation,
\[
Q_{abs}^{sphere} = 4 x {\rm Im}\left[\frac{\epsilon-1}{\epsilon+2}\right],
\]
where $x=2\pi a/\lambda$, $\epsilon=m^2$, and $m$ is the (wavelength-dependent)
complex index of refraction. This equation
combines the real and imaginary parts of the index of refraction
in a very specific, nonlinear way.
The peaks of silicate features from spherical particles are significantly 
shifted from other shapes \citep{yanamandra}.
To generalize the theoretical calculations, we fitted the shape
of the silicate feature using the `continuous distribution of ellipsoids'
(CDE) approximation, in which all possible axis ratios have equal
probability; the absorption efficiency averaged over shapes is then
\[
\langle Q_{abs}^{CDE}\rangle = \frac{8}{3} x {\rm Im}\left[
\frac{\epsilon}{\epsilon-1}\log\epsilon \right]
\]
which does not shift emission features from resonances as much, though 
it does tend to broaden them \citep{bh}.
This equation only applies in the small-particle limit, and the full
equations for a distribution of larger ellipsoids are not known and
probably computationally very challenging. 
Since the particles
producing the continuum of zodiacal light are large (as discussed above),
we cannot at present use a full theoretical CDE calculation (solving for
the temperature for each particle size) for the zodiacal
light. 
More sophisticated models of non-spherical particles, such as the
discrete dipole approximation \citep{dda} or T-matrix method \citep{tmatrix}
would be able to treat the problem self-consistently, but they are far
beyond the scope of this project.
However, the zodiacal light silicate {\it feature} is produced
by particles small enough that the small-particle limit can be used.
Therefore, we continue to use the models for spheres, as presented in the previous
sections, to calculate the continuum spectrum. To model the
silicate feature, we calculated CDE absorption efficiencies for all of
the materials and for 0.4 $\mu$m particle radii (see Fig.~\ref{emiss_size}).
A linear regression of the absorption efficiencies was fitted to the observed
spectrum. If the various silicate minerals have the same size distribution,
then the scale factor from the regression analysis gives the relative
abundance of each mineral.

The best fit to the silicate feature includes three components; 
Figure~\ref{plot_mix_cde} shows the mixture and each component.
This solution was found by trying each three-way combination of 40 different
theoretical models (each of the materials described above, some with
and without different amounts of carbonaceous material), and identifying
the solution with minimum $\chi^2$.
(Solutions with negative coefficients were discarded as nonphysical,
leaving 3741 models involving combinations of various materials.)
The most abundant silicate is amorphous olivine (with a scale factor of
1.12), which is the best single-mineral fit to the spectrum. 
However, amorphous olivine cannot fit the blue or red wing of the 
observed feature, which is distinctly `boxy.' 
The blue wing requires a different silicate mineral, and a reasonable
fit can be obtained with a hydrous silicate (montmorillonite, with a
scale factor of 0.25). There is some indication of a more detailed
substructure in the blue wing that can be fitted with this mineral, 
with a possible hump in the spectrum around 8.6 $\mu$m. None of the
minerals we considered have this structure, but it is relatively
subtle and uncertain so we did not pursue it further. 
The red wing of the zodiacal feature, including the
small peak at 11.35 $\mu$m, is reasonably fit by crystalline olivine (with
a scale factor of 0.12). 

\epsscale{0.7}
\plotone{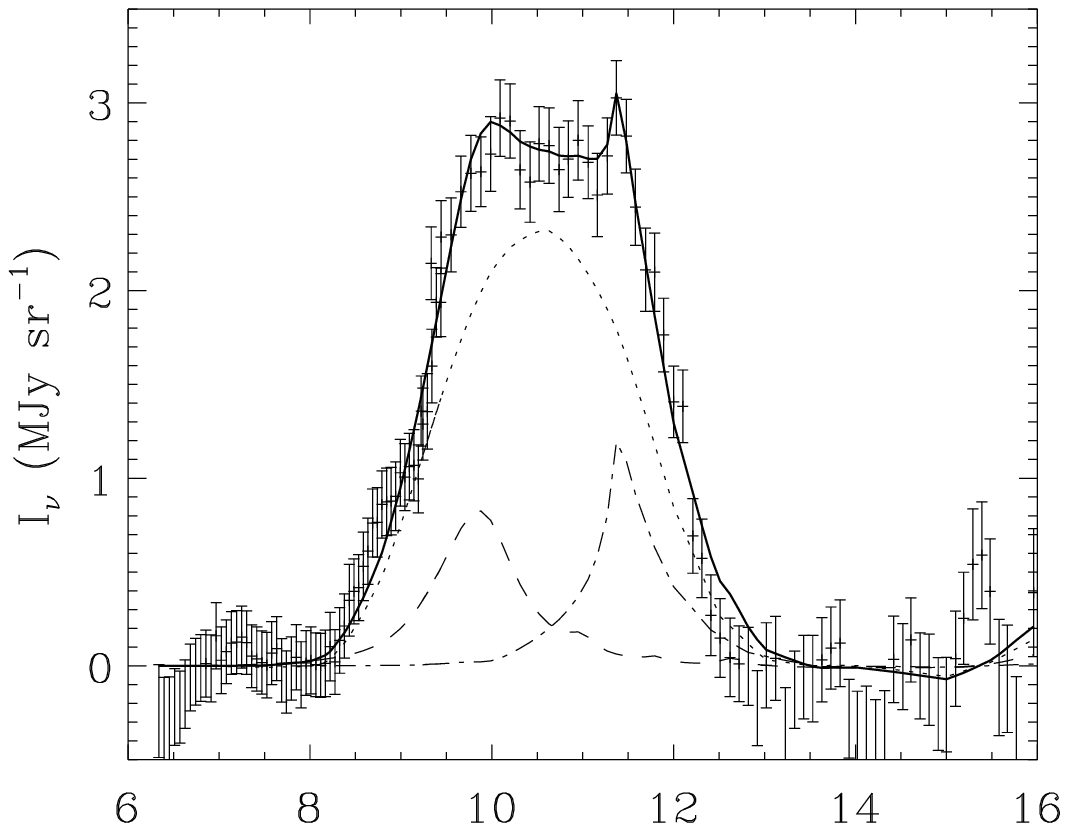}
\epsscale{1}
\figcaption{Comparison of a mixture of silicates with all possible
ellipsoidal shapes to the observed, continuum-subtracted, spectrum
of the zodiacal light. The solid line is the mixture fit, the
dotted line is the amorphous olivine contribution, the dashed line
is the montmorillonite contribution, and the dash-dotted line is the
crystalline olivine contribution.
\label{plot_mix_cde}}
\raggedright

The presence of a hydrous silicate in the mixture required to fit the shape
of the silicate feature indicates that the parent bodies of the zodiacal
particles were formed in a location where water could be mixed with
crystallizing minerals. This may suggest formation in the inner Solar
System, where temperatures and pressures were high, or processing on
an active surface (or in collisions). A mixture of water and minerals
is certainly expected for comets, but the water is less likely to be 
contained so deeply within the silicate minerals. These subtle clues
can reveal relationships between zodiacal particles and their parent bodies,
and they need further exploration.


Using the best-fitting model, we can predict the spectrum at wavelengths outside
the region we observed. Due to the prevalence of large particles, and the weakness
of the 9--11 $\mu$m feature, we predict
a relatively featureless spectrum. 
The CDE fit includes crystalline olivine, which has features
at {\it 20}, 21, {\it 24}, and {\it 35} $\mu$m,
and montmorillonite, which has features at 19, {\it 23}, and 29.5 $\mu$m
(brightest features in {\it italics}).
  We note that earlier measurements of montmorillonite  
  \citep{early_montmorillonite} 
  also show a broad feature in the 45 to 55 $\mu$m range that is not reproduced
  in their later derivations of the optical constants (which we use here) from Lorentz
  oscillators.
The amplitude of these features is not accurately predicted by the CDE
model, because it is a single-size model. A rough estimate can be made
by using the CDE mixture model to give the detailed features and 
the spherical model to give the continuum level. 
Interestingly, the strongest crystalline olivine and hydrous silicate features
alternate in wavelength, such that the sum is more like a lumpy continuum.
Only the 9--11 $\mu$m feature is strong in the mixture. The next
brightest features (in {\it italics} above) are only $\sim 1$ \% of
the continuum.
Thus, apart from the 9--11 $\mu$m feature, we do not expect significant structure
in the zodiacal light spectra obtained with the Infrared Spectrograph on
the {\it Space Infrared Telescope Facility} \citep{houck}.
By extension, we predict a relatively featureless spectrum for the terrestrial
portion of debris disks around mature stars, where the primordial dust has been 
removed and the particles are derived from asteroid collisions and 
comet fragmentation.

\def\extra{ 
Crystalline olivine
has features at 10.1, 11.0, 16.7,
18.8, 21.3, 27.5, and 32.5 $\mu$m; the amplitudes of the features relative to
their adjacent continuum are approximately 13, 16, 4.5, 3, 1.5, 6, and 1.4\%,
respectively. 
After multiplying by the fraction (10\%) of the mid-infrared emission due to this
material in the mixture model, no features brighter than 0.6\% (at 27.5 $\mu$m)
are predicted (other than those in the 9--11 $\mu$m range).
Amorphous forsterite has a feature at 18.8 $\mu$m, which roughly coincides
with the crystalline olivine feature; in the mixture, the spectral line is
predicted to be about 0.6\%.
}

\section{Relation between zodiacal light and parent bodies}

\subsection{Comparison to IDPs and cometary dust}

\citet{sandford} summarized the properties of interplanetary dust particles
(IDPs) collected in the atmosphere. He divided the particles into three classes:
olivines, pyroxenes, and layer-lattice silicates. 
All of the IDPs are heterogeneous, with the
carbonaceous material that survived atmospheric entry spread throughout the
grains. The collected particles are rather porous, and the `empty' regions
were probably at least partially filled with more volatile materials before
atmospheric entry. These properties are consistent with our `dirtied' silicate
sphere models, although the actual porosity of the grains in space is not
known and could affect the spectral properties. The infrared transmission
of a sample of IDPs was measured 
by \citet{sandfordwalker}. All of the silicates had 9--11 $\mu$m features.
The olivine particles presented substructures at 9.3, 10.1, and 11.2 $\mu$m,
while the pyroxenes presented substructures within the 9--11 $\mu$m band
that varied strongly from particle to particle. The layer-lattice
silicates have strong O--H features that would have been 
observable in the zodiacal spectrum. Thus, small particles composed of
layer-lattice silicates are unlikely to be very abundant in interplanetary
space.
The observed zodiacal light spectrum, on the other hand, bears a strong
resemblance to the olivine and pyroxene IDPs. 
The strong particle-to-particle mineralogy variations would, 
when blended together along the line
of sight in a zodiacal light observation, lead to a relatively broad and
smooth feature, which is in fact what we observe. 

Comets in the inner Solar System present relatively strong 9--11 $\mu$m 
silicate features \citep{hanner94}. These features vary from comet
to comet, and they may be stronger for new comets. The best-observed 
cometary silicate feature is for comet Hale-Bopp,
both because it was very bright and because of advances in infrared detection.
\citet{wooden} presented extensive ground-based observations of
the 9--11 $\mu$m Hale-Bopp, and \citet{crovisier} presented the {\it ISO} 
spectrum over a wider wavelength range. 
Comparing the continuum-subtracted zodiacal light
and Hale-Bopp silicate features (shown together in
Fig.~\ref{betapic_sil} below), the amplitude of the
silicate feature from the Hale-Bopp dust is twice the brightness of the
underlying continuum. In our nomenclature, $\Delta_{sil}$ ranged from
160 to 200 as the Hale-Bopp moved from 2.8 to 0.9 AU from the Sun
\citep{wooden}. Comets Halley and Mueller showed similarly strong
silicate features, while comets Borrelly and Faye had weaker features,
with $\Delta_{sil}\sim 20$ \citep{hanner94}.
The cometary silicate features are several to 20
times stronger than the silicate feature in the zodiacal light.
For comets with infrared spectra, the particles are inferred to
be much smaller than those that dominate the zodiacal light,
due both to their bright silicate features and
their continuum temperatures being much higher than blackbodies at
the corresponding distance from the Sun.
These coma grains are so small that they feel 
enough radiation pressure to put them in hyperbolic orbits. 
It is clear that these small dust grains are not the 
origin of interplanetary dust, unless the grains can grow
in the interplanetary medium.
\citet{molster} suggested that some Fe-poor crystalline
silicates may have condensed out of the gas phase close to the Sun, 
and the small grains transported by radiation pressure {\it over} the solar 
disk to distances beyond Jupiter.
Larger cometary particles have been detected by {\it in situ} observations
at P/Halley \citep{mcdonnell_halley}
and inferred by dynamic modeling of the morphology of the dust
around C/Austin \citep{lisse} and P/Encke \citep{reachencke}.
Let us assume that the small cometary grains have similar mineralogy to
the larger ones, so it is
useful to compare the mineralogy of the cometary and zodiacal particles. 
First, comet grains appear
to be more crystalline, with sharper substructures in their 9--11 $\mu$m
feature \citep{wooden}, and strong crystalline features in the 35 $\mu$m
region \citep{crovisier}. 
In Figure~\ref{betapic_sil}, the 11.3 $\mu$m crystalline olivine peak
is clear in the Hale-Bopp silicate feature.
Second, the comet grains appear to be Mg-rich
\citep{wooden}, which is similar to our results for the zodiacal
light, where a forsterite
material (Mg-rich olivine) matches the spectrum better than fayalite 
(Fe-rich).  We cannot say with confidence whether there is a real
distinction between what seem to be olivines (SiO$_4$) matching the
zodiacal spectrum and the mixture of olivines and 
pyroxenes (SiO$_3$) matching the Hale-Bopp
spectra \citep{wooden}, because the zodiacal spectrum is too dominated
by large particles to present clearly-defined spectral features.

\subsection{Spatial variations of the zodiacal spectrum:
cometary {\it versus} asteroidal origin for dust\label{sec_fit}}

A goal of this project was to determine how the zodiacal spectrum
varies from place to place. To first order, the spectra are remarkably
similar. The brightness is clearly higher at low ecliptic latitudes and
toward the Sun, as shown in the definitive maps of the zodiacal cloud 
by {\it COBE}/DIRBE \citep{kelsall}. 
For a more detailed comparison, Figure~\ref{figlutzfit} shows
the temperature and silicate excess determined from fits to the ISOCAM CVF
spectra described earlier in this paper. The fit values are displayed in 
the relevant Solar System coordinate system; specifically, they are plotted
as a function of angular separation from the Sun and absolute ecliptic
latitude. The spectra with the lowest color temperature are those taken
farthest from the Sun and at relatively lower ecliptic latitude. The warmest
spectra are those closest to the Sun and at high ecliptic latitude. 
These variations are intuitively expected, because the sight-lines that contain 
the coldest dust will be those that see farthest toward the outer Solar System.
The spectra taken closest to the Sun sample the hottest dust (reaching
as close as 0.9 AU). The reason the polar spectrum is among the warmest is
that it contains almost exclusively dust at 1 AU, while the ecliptic plane
spectrum closest to the Sun samples dust at a range of distances, from 0.9 AU
outward to distances where the dust is cooler. 

\plotone{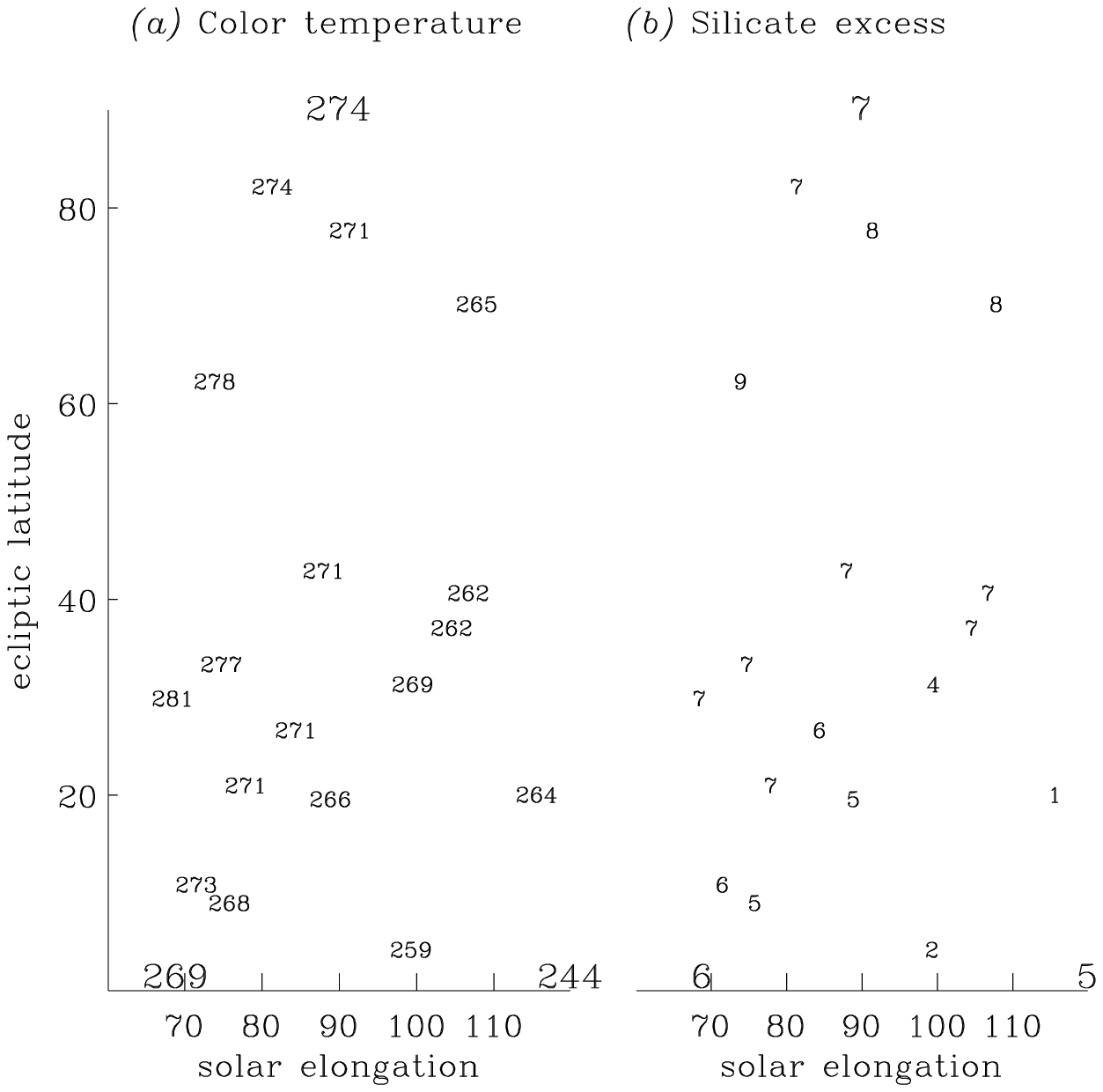}
\figcaption{The color temperature {\it (a)} and silicate excess
{\it (b)} derived from ISOCAM spectra of the zodiacal light. Both
the archival and dedicated zodiacal light observations are included;
the dedicated observations are distinguished by larger font.
\label{figlutzfit}}
\raggedright

The value of the polar temperature
is very similar to the temperature (278 K)
expected for rapidly-rotating, grey grains
at 1 AU from the Sun. This clearly demonstrates that the spectrum is dominated
by grains large enough that they emit efficiently in the mid-infrared. 
Specifically, $2\pi a/\lambda > 1$, where $\lambda\sim 30$ $\mu$m is the 
longest wavelength where the grains must emit efficiently in order to
behave as greybodies. This sets a limit on the grain radii of $a>5$ $\mu$m,
consistent with the theoretical models described above, which have the emission
produced by particles in the 10--100 $\mu$m size range.

The variation of the color temperature with viewing direction is qualitatively
as predicted by the DIRBE zodiacal light model \citep{kelsall}. 
To test this quantitatively, we evaluated a generalized version of
the DIRBE zodiacal light model (\S\ref{secdark}) over the mid-infrared
wavelengths at which the ISOCAM observations were made, and we used the
same color-temperature-fitting routine as we used on the data. 
The model yields colors with approximately the same variation of temperatures
with solar elongation and latitude. The pole-to-plane 
temperature difference at solar elongation $\epsilon=113^\circ$ is 30 K 
from ISOCAM and 24 K from the DIRBE model;
the in-plane ($\epsilon=66^\circ-113^\circ$) difference is 25 K from 
ISOCAM and 23 K from the DIRBE model. But the DIRBE model
is consistently warmer than the ISOCAM spectra, by $\sim 6$ K. 
The discrepancy can be traced to the high dust temperature at 1 AU,
$T_0=286$~K, used in the DIRBE model. 
This temperature used in the DIRBE model but was not included as a free
parameter; an independent, DIRBE-based model by \citet{wright98} also
adopted a value of $T_0$, but the adopted value was $T_0=280$ K, which
is in better agreement with the ISOCAM data.

While the continuum temperature was expected to show a reflex variation
as a function of viewing direction, due to changing dust temperature as
a function of distance from the Sun, the amplitude of the silicate feature,
expressed as a fraction of the continuum, is expected to remain relatively
constant within the inner Solar System unless the properties of the material
change from place to place. In Figure~\ref{figlutzfit}, the silicate 
feature amplitude shows a possible trend as a function of viewing direction.
Specifically, the feature is somewhat stronger at lower elongations and 
higher latitudes. This suggests that dust relatively farther from the
Sun has a weaker silicate feature than dust within $\sim 1.1$ AU of the
Sun, and possibly also dust with higher-inclination orbits has a somewhat
stronger silicate feature.

The variation of the silicate feature is small, if present, which suggests
that the mix of dust grains does not vary dramatically as a function of
distance from the Sun or from the ecliptic. This relative uniformity
applies over the range of distances within which the zodiacal light is 
produced. The radial range is not great: at 12 $\mu$m, in the ecliptic, toward 
$90^\circ$ elongation, 90\% of the zodiacal light is produced between the 
orbits of Earth and Mars \citep{reach91}.

A variation in the apparent mineralogy of dust in the coma of C/Hale-Bopp 
was detected by \citet{wooden}, who observed the silicate feature from
the comet at a range of heliocentric distances. They found that both
the amplitude of the feature (relative to the continuum) and its shape
varied with heliocentric distance. They inferred that there were
multiple silicate components (similar to our three-component fit to the 
zodiacal spectrum), and that the temperature of at least one of the
components was distinct from the others. The coldest component does
not become bright enough to contribute to the 9--11 $\mu$m feature
until it is relatively closer to the Sun. The net effect is that the
line-to-continuum ratio of the silicate feature increases as the 
heliocentric distance decreases. This effect goes in the same sense
as the effect seen in the zodiacal light, suggesting a similar
mechanism may be at play.

There have been other suggestions of variations of dust properties as a function
of location in the Solar System. \citet{renard} found that the dust 
albedo decreases as a function of distance from the Sun as
$A\propto r^{-0.32}$. That is, they found that grains
are more dark at greater distances from the Sun. 
This result could be qualitatively in accord with a decrease in the amplitude
of the silicate feature, if the grains are larger at greater distances
from the Sun, because larger grains have less silicate feature and
are relatively darker. (Larger grains are more efficient at infrared
absorption per unit visible absorption due to physical optics.)
Whether the albedo actually varies is extremely difficult to say with
confidence, and the same applies to our suggested variation in the silicate
feature. From remote sensing, we only retrieve an integral along the
line of sight, which cannot be uniquely inverted. The results from
the {\it Helios} spaceprobe, which measured the brightness, color, and
polarization of the zodiacal light while orbiting between 0.3 and 1 AU 
from the Sun, are considerably more reliable. \citet{leinert81} showed
that the {\it Helios} results require a radial variation of the 
volumetric scattering cross-section $nA\propto r^{-1.3\pm 0.05}$.
The \citet{kelsall} fit to the DIRBE infrared data yielded a density 
profile $n\propto r^{-1.34\pm.02}$ that is consistent with the volumetric 
scattering cross-section from {\it Helios}; 
combining these results, no variation of the albedo is required.
However, the infrared model assumed constant dust
properties, and it has not been determined whether a model with variable
dust properties fits the data significantly better.
The only convincing evidence for a variation in dust properties is the
{\it Helios} observation that dust closer
to the Sun generates relatively less polarization than dust at 1 AU
\citep{leinert81}.

Radial variations in dust properties, if present, can be explained
theoretically. The observations have suggested that dust is somewhat
darker, more polarizing, and less silicate-feature-producing at larger
distances from the Sun.
\citet{greenberghage} predicted varying properties of cometary dust,
with relatively larger number of small grains in the inner Solar System
being due to disintegration of cometary particles that consist
of many small grains stuck together by volatiles, which evaporate
as the grains approach the Sun. The same qualitative result is produced
by grain-grain collisions, regardless of particle morphology, as very large
grains, spiraling inward from the asteroid belt (for example) under
the influence of Poynting-Robertson drag, are
shattered by collisions before they reach the inner Solar System.
\citet{grun85} calculated the collisional lifetimes to be
shorter than the Poynting-Robertson lifetimes for grains larger than
100 $\mu$m. Asteroidal grains of this size would not survive into
the inner Solar System.

The carriers of the silicate feature are actually very small particles.
Figure~\ref{emiss_size} showed that the feature is produced most efficiently
by submicron particles. Such are predicted to have very short
lifetimes, because solar radiation pressure exceeds gravity
\citep{burns}. For the materials used in our mixture model, we
calculated the ratio of radiation pressure to gravitational
force, $\beta$. If the particles are compact, with volume densities
$\rho\sim 2$ g~cm$^{-3}$ such as typical of terrestrial rocks
and stratospheric-collected IDPs \citep{love},
then radiation pressure exceeds gravity ($\beta>1$) for 
silicate particles (both crystalline and amorphous)
with radius smaller than $\sim 0.4$ $\mu$m. 
(Amorphous carbon particles always have $\beta<1$, unless their densities 
are low.)
If the particles are porous,
then radiation pressure exceeds gravity for a proportionally larger radius,
because $\beta\propto 1/(\rho a)$.
Such particles should be relatively
rapidly ejected from the Solar System. However, it is clear that
particles of this size are present in the inner Solar System, as they were
observed by dust detectors aboard Pioneers 8 and 9 and HEOS-2 \citep{grun85}.
There is a continuous source of these so-called $\beta$-meteoroids, 
from the collisional comminution of larger particles that are spiraling toward the Sun
under the influence of Poynting-Robertson drag. 
Thus both comets
and asteroids particles can be sources of both large and small
particles in the inner Solar System: Poynting-Robertson drag brings
them in, and collisions grind them down.

To distinguish a cometary and asteroidal origin for the
zodiacal dust, a telling variation could be a dependence 
of the strength (or shape) of the silicate feature on ecliptic latitude. 
While asteroids are confined to the ecliptic,
comets have a wider range. 
New comets arrive at essentially all inclinations; however, 
nearly all of the dust they produce leaves the Solar System
due to the high eccentricity of the comet's orbit and the effects
of radiation pressure \citep{burns}. Short-period comets are the 
most plausible cometary source of zodiacal dust.
The mean inclination of main-belt
asteroid orbits is $7.9^\circ$, while the mean inclination of
comet orbits is $19^\circ$ \citep{allen}.
Thus dust from short-period comets would be on relatively higher-inclination
orbits than asteroidal dust, and it would be relatively more prevalent 
toward higher ecliptic latitudes. 

Can we see systematic differences
in the silicate feature versus ecliptic latitude?
In Figure~\ref{figlutzfit}, there is no strong trend.
The high-latitude spectra do show a silicate feature, and that feature
appears to have approximately average amplitude, though
perhaps on the high side.
We suspect that any dependence on the orbital inclination of the parent
body is masked by radial variations, so that from our vantage point
near the ecliptic plane we cannot easily separate radial from
vertical variations. The best chance for such a separation is to
consider the spectra along the meridian where the solar elongation is
90$^\circ$, and looking only at latitudes above $10^\circ$. 
Although the ecliptic plane will clearly contain
relatively more distant dust, by moving along this meridian we will
keep the mean solar distance to the emitting region at least roughly
constant. We find that the high latitude points have an silicate excess
$\Delta_{sil}=7\pm 1$, while the low-latitude points have 
$\Delta_{sil}=7\pm 1$. Thus there is no evidence for a segregation of
dust properties as a function of ecliptic latitude.
Our result (or non-result) is still consistent with 
estimates of parent sources for the interplanetary dust, which
show that both comets and asteroids are viable sources.
For example, \citet{whipple} predicted that short-period comets produce
enough dust to maintain the zodiacal cloud, while \citet{grogan} showed
that the asteroid families contribute a third of the cloud and
non-family asteroids can produce the rest.

\section{Comparison of zodiacal and exozodiacal spectra\label{exosec}}

The new mid-infrared spectra obtained from {\it ISO} allow us to compare
the properties of material in the interplanetary medium, comets, and
in debris disks around other stars. 
Figure~\ref{betapic_sil} compares the spectrum of the zodiacal light,
comet Hale-Bopp \citep{crovisier}, the $\beta$ Pic debris disk,
and the interstellar medium \citep{dl84}. To make this comparison, we have
subtracted smooth continua from each spectrum. The zodiacal light continuum
was obtained by first removing the \citet{kelsall} zodiacal light 
prediction for this line of sight and observing date.
(This model uses a blackbody kernel so it will not affect the 
silicate feature shape).
Then we subtracted a second-order polynomial in the 5--16 $\mu$m range
excluding 8--12 $\mu$m. For Hale-Bopp, the continuum is simply a
second-order polynomial fitted over this same wavelength range.
For the interstellar spectrum, we used the theoretical model described
above for astronomical silicate particles heated by the Sun at 1 AU 
with the interstellar size distribution \citep{mrn}, and then subtracted a 
second-order polynomial fitted over the same wavelength range as
for the other spectra.

\epsscale{0.85}
\plotone{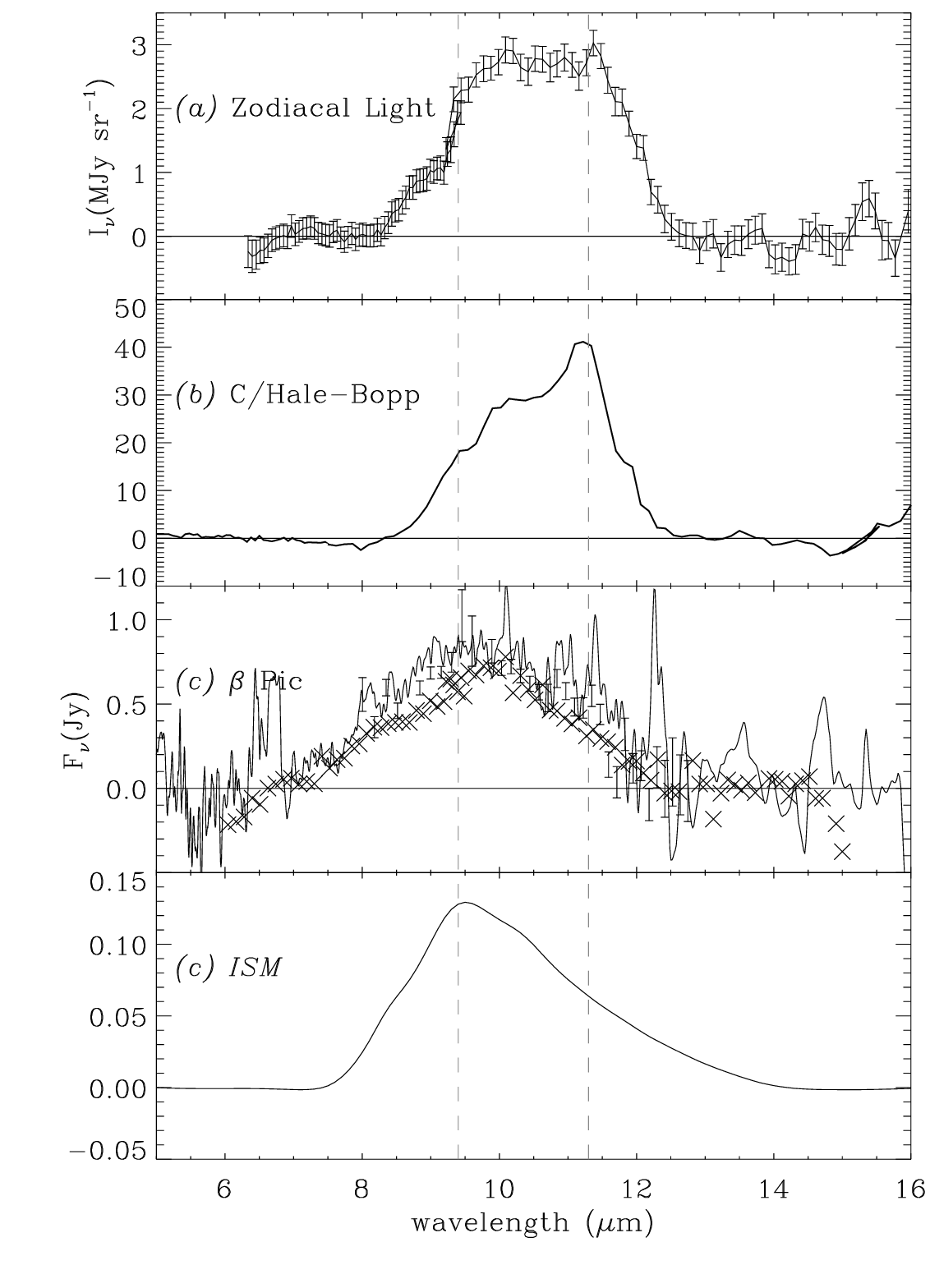}
\epsscale{1}
\figcaption{
Continuum-subtracted spectra of {\it (a)} the zodiacal light, 
{\it (b)} comet Hale-Bopp \citep{crovisier}, 
{\it (c)} the $\beta$ Pic debris disk, and 
{\it (d)} interstellar particles \citep{dl84}.
The spectra are roughly in evolutionary order, with the initial building blocks
from the interstellar medium at the bottom, then the agglomerating particles
in the $\beta$ Pic disk, then the comet, then the cometary and asteroidal
fragments of the zodiacal light.
In panel {\it (c)}, the observed spectra (same curve and symbol coding
as Fig.~\ref{betapic}) minus the photosphere and continuum debris disk
are plotted.
Comparing the zodiacal and exozodiacal features, the former
 appears redder and relatively boxy (like the 
cometary spectrum), while the $\beta$ Pic feature appears 
smoother and somewhat triangular.
Vertical dashed lines are shown at two suggestive wavelengths: the 11.3 $\mu$m
peak in the cometary spectrum (attributed to olivine) and the 9.4 $\mu$m
peak of the interstellar astronomical silicate spectrum. 
(In panel {\it (c)}, the
emission features in the 6.4--6.8 $\mu$m range are due to H$_2$O.)
\label{betapic_sil}}
\raggedright

For $\beta$ Pic, the continuum was modeled using a stellar atmosphere model
for the star and a simple approximation of the disk.
The separation of the silicate feature and continuum is 
difficult for $\beta$ Pic because the feature is so smooth, and
there is such a wide range of physical conditions contributing along
the line of sight---from the outer disk, which dominates at long wavelengths
\citep{heinrichsen},
to the inner disk, which is only revealed at mid-infrared wavelengths 
\citep{pantin} and high-resolution, coronagraphic  optical images \citep{heap}.
For the large-aperture observations, which include the entire debris disk,
we made a simultaneous fit to the 
continuum in the SWS and PHOT data (5--200 $\mu$m) using two 
modified blackbody components:
\begin{equation}
F_\nu = [(1.07\pm 0.11)\times 10^{-5} B_\nu(77.5\pm 1.4\,\,{\rm K}) + 
(2.37\pm 0.18\times 10^{-8}) B_\nu(197.5\pm 2.5\,\,{\rm K})] (12 \mu{\rm m}/\lambda)
\label{bpsws}
\end{equation}
where $B_\nu(T)$ is the Planck function. The two components represent
the `outer' and `inner' exozodiacal light, respectively,
with the `outer' component
being the one that dominates the far-infrared brightness and the 
mass of the disk. 
This fit, shown as a dashed line in Figure~\ref{betapic},
becomes brighter than the photosphere at wavelengths longer than 12 $\mu$m. 
For the smaller aperture observations by ISOCAM, we approximated
the continuum with a single blackbody component (hence the fit is only
valid at wavelengths shorter than 15 $\mu$m),
\begin{equation}
F_\nu = 1.54 \times 10^{-9} B_\nu(300\,\,{\rm K}),
\label{bpcam}
\end{equation}
which is shown as a dotted line in Figure~\ref{betapic}. The continuum
is weaker in the ISOCAM and ground-based mid-infrared spectra,
compared to the SWS and PHOT data, because part 
of the `outer' disk is resolved and excluded from the former.
The comparable temperatures of the ISOCAM continuum and the `inner' component
of the SWS/PHOT continuum fit suggest they both arise from similar parts of
the disk, but even the `inner' SWS/PHOT continuum is produced from a somewhat
greater distance from the star than the CAM/ground-based continuum.
Recent high-angular-resolution observations with Keck 
by \citet{weinberger} resolve the thermal emission from
the disk even better than those described here,
revealing that the spectral shape changes
with distance from the star, such that the silicate feature is
largely produced from the inner $1''$ of the disk.
A more detailed model for the debris disk is needed to study the shape of the
continuum feature in detail; for our purposes, the simple models
are sufficient to allow us to extract the silicate feature.

\def\extra{
The IRTF observations were made with a 3.7'' diameter aperture. 
The {\it ISO} observations, both spectroscopic and photometric,
used much larger apertures that include the entire disk. 
In Figure~\ref{betapic_sil} it is clear that 
the {\it ISO} points are consistently higher than the IRTF points, suggesting
that the IRTF spectrum does not include the entire disk. 
(Recall 1''=19 AU at the distance of $\beta$ Pic.)
The {\it ISO} spectrum is dominated by the colder out disk and
contains, the IRTF spectrum is relatively more sensitive to
the inner disk, and the Keck spectrum isolates the inner disk.
}

The $\beta$ Pic silicate feature appears similar to that found
in recently-published exozodiacal light studies. 
After the first detection of the silicate feature from $\beta$ Pic,
\citet{telknack} had already noted that
the silicate feature is wider than that the interstellar feature \citep{dl84}.
\citet{waelkens} showed that the silicate features of $\beta$ Pic and 51 Oph 
are very similar. 
\citet{vandenancker}
presented the remarkable spectrum of 51 Oph, for which the debris disk
is so bright that subtraction of the central star is hardly needed in
the mid-infrared. They noted the extremely wide, and somewhat triangular,
shape of the silicate feature, and hypothesized that the red wing of the 
feature, which is wider than silicate minerals we have studied to date,
may be due to a blend with a second mineral. They also found that 
the 51 Oph silicate feature is identical to that of symbiotic star
V385 Cen but wider (on the red side) than that of Herbig Ae star AB Aur,
leading to the tantalizing speculation that the dust in 51 Oph may 
be `` `fresh', {\it i.e.} recently formed rather than accreted from the
interstellar medium and processed in a proto-planetary disk''
\citep{vandenancker}. 
Given the extreme similarity of silicate features between $\beta$ Pic and
51 Oph, it is likely that the broad, triangular shape may be a
characteristic feature of proto-planetary disks, and may
in fact be due to processing (including agglomeration) occurring in the disk.

It is clear that the overall spectrum of the zodiacal light is very different
from exozodiacal light as evidenced by $\beta$ Pic or 51 Oph.
These exozodiacal spectra are dominated by dust much further from their
exciting stars and therefore much colder. The 10 $\mu$m silicate feature from
of these disks is produced by dust in the inner disks. Thus the
line-to-continuum ratio (which is a good indicator of particle size
and mineralogy) is very difficult to compare: one would have to
isolate the contributions form the outer and inner disk. Using the
large-aperture exozodiacal spectra, we can still compare the shapes
of the silicate features:
the shape of the silicate feature from $\beta$ Pic is quite
different from that of comet Hale-Bopp, and it is somewhat different
from that of the zodiacal light.
While the zodiacal feature is relatively
well-defined and boxy, the integrated exozodiacal feature is more broad and 
triangular. The cometary feature has a pronounced 11.3 $\mu$m peak, which
is not present in the zodiacal or exozodiacal spectra.
Silicates producing the 11.3 $\mu$m feature are likely present
in both exozodiacal disks; they explain the boxy shape of the red side of the feature. 
In $\beta$ Pic, and 51 Oph, no substructure is evident in the
integrated spectrum of the entire disk, at the limit of 
{\it ISO} sensitivity. 
The high-angular-resolution spectra \citep{knacke,weinberger} of the inner portion
of the $\beta$ Pic disk are actually similar to that of the zodiacal light.
Any differences between the zodiacal and
integrated exozodiacal may be due to different mineralogy,
radiation environment, or mineralogy
and temperatures contributing to the total flux.
Indeed, the particle sizes and properties should be very different in
the younger systems. While the younger systems retain their protoplanetary 
disks of dust and gas, the Solar System particles are produced by crumbling of asteroids
and comets. The presence of H$_2$O, CO, and CO$_2$ disks around both 51 Oph
\citep{vandenancker} and $\beta$ Pic (Fig.~\ref{betapic}) may lead
to a significant gas-grain chemistry that changes particle properties.
Also, while the young systems have very dense disks completely dominated
by collisions, the Solar System is rather rarefied, and the dynamics
of particles that produce the zodiacal light are dominated by Poynting-Robertson
drag.

\section{Conclusions}

We have presented and contrasted the mid-infrared (5--16 $\mu$m) spectra of 
the zodiacal light and the exozodiacal light around $\beta$ Pic. 
These two phenomena are due to planetary systems in very different stages of
evolution: while the $\beta$ Pic is likely in an early, particle-agglomeration
stage leading to planet formation, the Solar System is in an advanced,
large-body-colliding stage. The resulting particle properties, as seen in their
mid-infrared spectra, are very different. For both systems, the mid-infrared
emission is due to a combination of continuum emission and a broad silicate
feature. 
The most glaring difference between $\beta$ Pic and the Solar System
is the presence of
significant cold dust $\simgt 30$ AU from $\beta$ Pic that dominates the
far-infrared emission and contributes significantly even in the mid-infrared.
The presence of a silicate feature in both zodiacal and exozodiacal spectra
clearly illustrates their differences. The zodiacal feature is weak,
at only 5--7 \% of the continuum, while the exozodiacal feature is 
brighter than the continuum.
The amplitude of the silicate feature is an indicator of particle size. 
Theoretical models showed that the spectrum is dominated by emission from 
particles with radii in the range 10--100 $\mu$m. The exozodiacal
feature, in contrast, is produced by smaller particles with
radii smaller than $\lambda/2\pi=2$ $\mu$m. 
The shape of the zodiacal silicate feature, similar to the feature seen 
in comets and interplanetary dust particles, spans the range 9--12 $\mu$m and 
is boxy. The exozodiacal ($\beta$ Pic and 51 Oph) feature is broader and 
more triangular. Neither the zodiacal nor the exozodiacal silicate features
are identical to the interstellar dust feature, which peaks on the blue side
and tails off on the red side. There is a suggestive progression in
the silicate feature shape as a function of evolutionary age: the 
initial interstellar building blocks transform into the
cometary spectrum with the $\beta$ Pic spectrum as an intermediate
step.


\acknowledgements 
This work is based on observations with ISO, an ESA project with instruments funded by ESA Member States
 (especially the PI countries: France, Germany, the Netherlands and the
United Kingdom) with the participation of ISAS and NASA. We gratefully acknowledge
help from Joris Blommaert with ISOCAM calibration issues. The search for
variations in the zodiacal spectrum was inspired by 
Alain L\'eger. The research described in this paper was carried out at
the California Institute of Technology under a contract with the 
National Aeronautics and Space Administration.


\clearpage

\begin{thebibliography}{} 
 

\def\pp{\parshape 2 0truecm 16truecm 1truecm 15truecm}

\def\refic #1;#2;#3;#4;#5;#6 {\noindent \pp{\sc #1} {#2}. {#3}. {\it
#4}. {\bf #5}, {#6}.}\par 

\def\refpre #1;#2;#3;#4 {\noindent \pp{\sc #1} {#2}. {#3}, {#4}.}\par

\def\refbook #1;#2;#3;#4;#5;#6;#7;#8 {\noindent \pp{\sc #1} {#2}. #3.
In {\it #4} ({#5}, Eds.), pp. #6. #7, #8.} 

\def\refbookpre #1;#2;#3;#4;#5;#6;#7;#8 {\noindent \pp{\sc #1} {#2}.
#3. In {\it #4} ({#5}, Eds.), #6. #7, #8.} 


\def\extra{
\bibitem[Backman {\it et al.}(1997)]{backman97} \refic Backman, D. E., 
S. B. Fajardo-Acosta, R. E. Stencel, and J. R. Stauffer; 1997; Dust Disks 
around Main Sequence Stars; \apss; 255; 91--101
}

\bibitem[Beckwith, Henning, and Nakagawa(2000)]{beckwith} \refbook Beckwith, S. V. W., T. Henning, and Y. Nakagawa; 2000; Dust
properties and assembly of large particles in protoplanetary 
disks;Protostars and Planets IV;V. Mannings, A. P. Boss, and
S. S. Russell;533--558;U. Arizona Press;Tucson

\bibitem[Biviano {\it et al.}(1998)]{bivianoflat} \refpre Biviano, A., J. Blommaert,
O. Laurent, K. Okumura, R. Siebenmorgen, B. Altieri, O. Boulade,
L. Metcalfe, \& S. Ott;1998;The ISOCAM Flat Field Calibration 
Report;European Space Agency

\bibitem[Blommaert {\it et al.}(2001)]{blommcvf} \refpre Blommaert, J. A. D. L.,
F. Boulanger, and K. Okumura;2001;ISOCAM CVF photometry report;ESA technical 
report 2001: SAI/2001-034/Rp, ESTEC, Noorwijk.

\bibitem[Bohren and Huffman(1983)]{bh} {\noindent \pp{\sc Bohren, C. F. 
and D. R. Huffman} 1983. {\it Absorption and Scattering of Light by Small 
Particles}. Wiley, New York.}

\bibitem[Boulanger {\it et al.}(1996)]{boulrhooph} \refic Boulanger, F. and
16 colleagues;1996;Mid-Infrared imaging and spectroscopy in 
Ophiuchus;\aap;315;L325--L328

\bibitem[Bradley {\it et al.}(1999)]{bradley99} \refic Bradley, J. P., 
L. P. Keller, T. P. Snow, M. S. Hanner, G. J. Flynn, J. C. Gezo, S. J. 
Clemett, D. E. Brownlee, and J. E. Bowey;1999;An infrared spectral match 
between GEMS and interstellar grains;Science;285;1716--1718

\bibitem[Burns, Lamy, and Soter(1979)]{burns} \refic Burns, J. A., P. L. Lamy,
and S. Soter;1979;Radiation 
forces on small particles in the Solar System;Icarus;40;1--48

\def\extra{
\bibitem[Ceplecha {\it et al.}(1998)]{ceplecha} \refic Ceplecha, Z. K., J. I. Borovicka, W. G. Elford, D. O. Revelle,
R. L. Hawkes, V. Porubcan, and M. Simek;1998;Meteor phenomena and 
bodies;Space Sci. Rev;84;327--471
}
		  
\bibitem[Cesarsky {\it et al.}(1996)]{cesarsky} \refic Cesarsky, C. J., A. Abergel, 
P. Agnese, B. Altieri, J. L. Augueres, H. Aussel, A. Biviano, J. Blommaert, 
J. F. Bonnal, F. Bortoletto, and 56 coauthors;1996;ISOCAM in flight;\aap;315;L32--L37

\bibitem[Coulais {\it et al.}(2000)]{coulais} \refic Coulais, A. and 
A. Abergel;2000;Transient correction of the LW-ISOCAM data for
low contrasted illumination;Astron. Astrophs. Suppl;141;533--544

\bibitem[Cox(2000)]{allen} {\noindent \pp{{\sc Cox, A. N.} 
{2000}. {\it Allen's Astrophysical Quantities}. Springer-Verlag, New York.}}

\bibitem[Crovisier {\it et al.}(1997)]{crovisier} \refic Crovisier, J., 
K. Leech, D. Bockelee-Morvan, T. Y. Brooke, M. S. Hanner, B. Altieri, H. U. 
Keller, and E. Lellouch;1997;The spectrum of Comet Hale-Bopp (C/1995 01) 
observed with the Infrared Space Observatory at 2.9 AU from the 
Sun;Science;275;1904--1907

\bibitem[de Graauw {\it et al.}(1996)]{degraauw} \refic de Graauw, T., and
60 co-authors;1996;Observing with the ISO Short-Wavelength 
Spectrometer;\aap;315;L49--L54

\bibitem[Demyk {\it et al.}(2001)]{demyk} \refic Demyk, K., Ph. Carrez, Ph., 
H. Leroux, P. Cordier, A. P. Jones, J. Borg, E. Quirico, P. I. Raynal, 
and L. d'Hendecourt;2001;Structural and chemical alteration of crystalline 
olivine under low energy He$^+$ irradiation;\aap;368;L38--L41

\bibitem[Dermott {\it et al.}(1984)]{dermottband} \refic Dermott, S. F., P. D. Nicholson, 
J. A. Burns, J. R. Houck;1984;Origin of the solar system dust bands 
discovered by IRAS;Nature;312;505--509

\bibitem[Dermott {\it et al.}(1994)]{dermottring} \refic Dermott, S. F., S. Jayaraman, 
Y. L. Xu, B. A. S. Gustafson, J. C. Liou, J. C.;1994;A circumsolar ring of asteroidal 
dust in resonant lock with the Earth;Nature;369;719--723

\bibitem[Dorschner {\it et al.}(1995)]{dorschner95} \refic Dorschner, J.,
B. Begemann, T. Henning, C. J\"ager, and H. Mutschke;1995;Steps
toward interstellar silicate mineralogy. II. Study of Mg-Fe-silicate
glasses of variable composition;\aap;300;503--520

\bibitem[Draine(1988)]{dda} \refic Draine, B. T.;1988;The discrete-dipole approximation 
and its application to interstellar graphite grains;\apj;333;848--872

\bibitem[Draine and Lee(1984)]{dl84} \refic Draine, B. T. 
and H. M. Lee;1984;Optical properties of interstellar graphite and 
silicate grains;\apj;285;89--108

\bibitem[Edoh(1983)]{edoh} \refpre Edoh, O.;1983;Optical Properties of Carbon 
from the Far Infrared to the Far Ultraviolet;Ph. D. thesis, University of Arizona

\bibitem[Fabian {\it et al.}(2001)]{fabian} \refic Fabian, D., T. Henning,
C. J\"ager, H. Mutschke, J. Dorschner, and O. Wehrhan;2001;Steps
toward interstellar silicate mineralogy. VI. Dependence of crystalline 
olivine IR spectra on iron content and particle shape;\aap;378;228--238

\bibitem[Fajardo-Acosta and Knacke(1995)]{fajardo95} \refic Fajardo-Acosta, S. B.
and R. F. Knacke;1995;IRAS low resolution spectra with $\beta$ Pictoris-type
silicate emission;\aap;295;767--774

\bibitem[Greenberg and Hage(1990)]{greenberghage} \refic Greenberg, J. M. and
J. I. Hage;1990;From interstellar dust to comets -- A unification of
observational constrants;\apj;361;260--274

\bibitem[Grogan, Dermott, and Durda(2001)]{grogan} \refic Grogan, K.,
S. F. Dermott, and D. D. Durda;2001;The size-frequency distribution of
the zodiacal cloud: Evidence from the Solar System dust 
bands;Icarus;152;251--267

\bibitem[Gr\"un {\it et al.}(1985)]{grun85} \refic Gr\"un, E., H. A. 
Zook, H. Fechtig, and R. H. Giese;1985;Collisional balance of the 
meteoritic complex;Icarus;62;244--272

\bibitem[Gr\"un {\it et al.}(2001)]{grun01} \refic Gr\"un, E., and 23 
colleagues;2001;Broadband infrared photometry of comet Hale-Bopp
with ISOPHOT;\aap;377;1098--1118

\bibitem[Hanner(1984)]{hanner84} \refic Hanner, M. S.;1984;A comparison
of the dust properties in recent periodic comets;Adv. Space Res.;4.9;189--196

\bibitem[Hanner, Lynch, and Russell(1994)]{hanner94} \refic Hanner, M. S.,
D. K. Lynch, and R. W. Russell;1994;The 8--13 micron spectra of comets
and the composition of silicate grains;\apj;425;274--285

\bibitem[Hayward, Hanner, and Sekanina(2000)]{hayward} \refic 
Hayward, T. L., M. S. Hanner, and Z. Sekanina;2000;Thermal Infrared 
Imaging and Spectroscopy of Comet Hale-Bopp (C/1995 O1);\apj;538;428--455

\bibitem[Heap {\it et al.}(2000)]{heap} \refic Heap, S. R.. D. J. Lindler,
T. M. Lanz, M., R. H. Cornett, I. Hubeny, S. P. Maran, and
B. Woodgate;2000;Space Telescope Imaging Spectrograph Coronagraphic 
Observations of $\beta$ Pic;\apj;539;435--444

\bibitem[Heinrichsen {\it et al.}(2000)]{heinrichsen} \refic 
Heinrichsen, I., H. J. Walker, U. Klaas, U., R. J. Sylvester, and
D. Lemke, D.;1999;An infrared image of the dust disc around beta 
PIC;\mnras;304;589--594

\bibitem[Houck(1995)]{houck} \refic Houck, J. R. and J. E. Van Cleve;1995;IRS: an infrared 
spectrograph for SIRTF;Proc. Soc. Photo-Opt. Inst. Eng;2475;456--463

\bibitem[J\"ager {\it et al.}(1998)]{jaeger98} \refic J\"ager, C., F. J. Molster,
J. Dorschner, Th. Henning, H. Mutschke, and L. B. F. M. Waters;1998;Steps 
toward interstellar silicate mineralogy. IV. The crystalline 
revolution;\aap;339;904--916

\bibitem[Kelsall {\it et al.}(1998)]{kelsall} \refic Kelsall,  T., J. L. Weiland, B. A. Franz, W. T. Reach,
	R. G. Arendt, E. Dwek, H. T. Freudenreich, M. G. Hauser, 
	S. H. Moseley, N. P. Odegard, R. F. Silverberg, and
	E. L. Wright;1998;The COBE Diffuse Infrared Background Experiment
	search for the cosmic infrared background. II. Model of the 
	interplanetary dust cloud;ApJ;508;44--73
	
\bibitem[Kessler {\it et al.}(1996)]{kessler} \refic Kessler, M. F., J. A. Steinz, 
M. E. Anderegg, J. Clavel, G. Drechsel, P. Estaria, J. Faelker, J. R. Riedinger, 
A. Robson, B. G. Taylor, S. Ximenez de Ferran;1996;The Infrared Space Observatory 
(ISO) mission;\aap;315;L27--L31

\bibitem[Knacke {\it et al.}(1993)]{knacke} \refic Knacke, R. F., S. 
B. Fajardo-Acosta, C. M. Telesco, J. A. Hackwell, D. K. Lynch, and R. W. 
Russell;1993;The Silicates in the Disk of beta Pictoris;\apj;418;440--450

\bibitem[Koike and Hiroshi(1990)]{montmorillonite} \refic Koike, C., and S. 
Hiroshi;1990;Optical Constants of Hydrous Silicates from 7 $\mu$m to 
400 $\mu$m;\mnras;246;332--336

\bibitem[Koike, Hasegawa, and Hattori(1982)]{early_montmorillonite} \refic Koike, C., 
H. Hasegawa, and T. Hattori;1982;Mid- and far-infrared extinction coefficients of hydrous 
silicate minerals;\apss;88;89--98

\bibitem[Lagage {\it et al.}(1999)]{lagage} \refbook Lagage, P.-O., O. Boulade,
C. J. Cesarsky, T. Douvion, V. Mannings, E. Pantin, and 
A. Sargent;1999;ISOCAM spectro-imaging observations of the $\beta$-Pictoris
dust disk;The Universe as seen by ISO;P. Cox and M. F. 
Kessler;207--210;ESA;Noordwijk

\bibitem[Lamy, Gr\"un, and Perrin(1987)]{lamy_halley} \refic Lamy, P. L., E. Gr\"un, and J. M. 
Perrin;1987;Comet P/Halley--implications
of the mass distribution function for the photopolarimetric properties of the 
dust coma;\aap;187;767--773

\bibitem[Landgraf {\it et al.}(2000)]{landgraf} \refic Landgraf, M., W. J. Baggaley, E. Gr\"un, H. Kr\"uger, and 
G. Linkert;2000;Aspects of the mass distribution of interstellar grains
in the Solar System from in-site measurements;\jgr;105;10343--10352

\bibitem[Laor and Draine(1993)]{laor}  \refic Laor, A. and B. T. 
Draine;1993;Spectroscopic constraints on the
properties of dust in active galactic nuclei;\apj;402;441--468

\bibitem[Leinert {\it et al.}(1981)]{leinert81} \refic Leinert, C., I. Richter, E. Pitz, 
and B. Planck;1981;The zodiacal light from 1.0 to 0.3 AU as observed by the Helios
space probes;\aap;103;177--188

\def\extra{
\bibitem[Liou and Zook(1998)]{liou} \refic Liou, J. C., and H. A. Zook;1996;Comets as a source of low
eccentricity and low inclination interplanetary dust
particles;Icarus;123;491--502
}

\bibitem[Lisse {\it et al.}(1994)]{lisse} \refic Lisse, C. M., H. T. Freudenreich, 
M. G. Hauser, T. Kelsall, S. H. Moseley, W. T. Reach, and 
R. Silverberg;1994;Infrared Observations of Comet
Austin (1990V) by the COBE/Diffuse Infrared Background 
Experiment;\apjl;432;L71--L74

\bibitem[Love, Joswiak, and Brownlee(1994)]{love} \refic Love, S. G., D. J. Joswiak,
and D. E. Brownlee;1994;Densities of stratospheric micrometeorites;Icarus;111;227--236

\bibitem[Malfait {\it et al.}(1999)]{malfait} \refic Malfait, K., C. 
Waelkens, J. Bouwman, A. de Koter, and L. B. F. M. Waters;1999;The ISO 
spectrum of the young star HD 142527;\aap;345;181--186

\bibitem[Mathis, Rumpl, and Nordsieck(1977)]{mrn} \refic Mathis, J. S.,
W. Rumpl, and K. H. Nordsieck;1977;The size distribution of 
interstellar grains;\apj;217;425--433

\bibitem[McDonnell {\it et al.}(1991)]{mcdonnell_halley} \refbook McDonnell, J. A. M.,
P. L. Lamy, and G. S. Pankiewicz;1991;Physical properties of cometary
dust;Comets in the post-Halley era;R. L. Newburn, Jr., M. Neugebauer, and
J. Rahe;1043--1073;Kluwer;Dordrecht

\bibitem[McDonnell and Gardner(1998)]{mcdonnell98} \refic McDonnell, J. A. M.
and D. J. Gardner;1998;Meteoroid morphology and densities: decoding
satellite impact data;Icarus;133;25--35

\bibitem[Meeus {\it et al.}(2001)]{muess} \refic Meeus, G., L. B. 
F. M. Waters, J. Bouwman, M. E. van den Ancker, C. Waelkens, and K. 
Malfait;2001;ISO spectroscopy of circumstellar dust in 14 Herbig Ae/Be 
systems: Towards an understanding of dust processing;\aap;365;476--490

\bibitem[Mischenko(1991)]{tmatrix} \refic Mischenko, M. I.;1991;Extinction and 
polarization of transmitted light by partially aligned nonspherical 
grains;\apj;367;561-574



\bibitem[Molster, Bradley, and Sitko(2002)]{molster} \refic Molster, F. J., 
J. P. Bradley, and M. L. Sitko;2002;The Origin of the Crystalline Silicates in our Solar 
System;Lunar Planet. Sci;XXXIII;33;1471

\bibitem[Mukai and Koike(1990)]{mukai90} \refic Mukai, T., and C. Koike;1990;Optical constants 
of olivine particles between
wavelengths of 7 and 200 microns;Icarus;87;180--187


\bibitem[Okumura(2001)]{koryostray} \refbook Okumura, K.;2001;Optical 
properties of ISOCAM and/or ISO telescope;The 
Calibration Legacy of the ISO mission (ESA SP-481);ESA;48;ESA;Noorwijk

\bibitem[Ootsubo(2002)]{ootsubothesis} \refpre Ootsubo, T.;2002;Study
on interplanetary dust based on the mid-infrared observation of
zodiacal emission by the IRTS;Thesis, University of Tokyo.

\bibitem[Ootsubo {\it et al.}(1998)]{ootsubo98} \refic Ootsubo, T., T. Onaka, 
I. Yamamura, T. Tanab\'e, T. L. Roellig, L.-W. Chan, and 
T. Matsumoto;1998;IRTS observation of the mid-infrared spectrum of the
zodiacal emission;Earth~Planets~Space;50;507--511


\bibitem[Pantin, Lagage, and Artymowicz(1997)]{pantin} \refic Pantin, E.,
P.-O. Lagage, and P. Artymowicz;1997;Mid-infrared images and models of 
the beta Pictoris dust disk;\aap;327;1123--1136

\bibitem[Pantin, Waelkens, and Malfait]{pantinwaelkens} \refbook Pantin, E.,
C. Waelkens, and K. Malfait;1999;SWS observations of the $\beta$ Pictoris
dust disk;The Universe as seen by ISO;P. Cox and M. F. 
Kessler;385--388;ESA;Noordwijk

\bibitem[Preibisch(1993)]{preibisch} \refic Preibisch, Th., V. Ossenkopf, H. W. York, and 
Th. Henning;1993;The influence
of ice-coated grains on protostellar spectra;\aap;279;577--588

\bibitem[Reach(1988)]{reach88} \refic Reach, W. T.;1988;
Zodiacal emission. I - Dust near the earth's orbit;\apj;335;468--485

\bibitem[Reach(1991)]{reach91} \refic Reach, W. T.;1991;Zodiacal Emission. II.
Dust near ecliptic;\apj;369;529--543

\bibitem[Reach(2000)]{reachmemo} \refpre Reach, W. T.;2000;SIRTF background
estimation: methods and implementation;{SIRTF Science Center Memo 
(http://sirtf.caltech.edu/SSC/C\_PropKit/bgdoc\_release.pdf): 
Caltech, Pasadena}

\bibitem[Reach {\it et al.}(1995)]{reachring} \refic Reach, W. T.,
B. A. Franz, J. L. Weiland, M. G. Hauser, T. N. Kelsall, E. L. Wright, 
G. Rawley, S. W. Stemwedel, W. J. Spiesman;1995;Observational Confirmation of 
a Circumsolar Dust Ring by the COBE Satellite;Nature;374;521--523

\bibitem[Reach {\it et al.}(1996)]{reach96} \refic Reach, W. T. and 
12 colleagues;1996;Mid-Infrared spectrum of the zodiacal light.;\aap;
315;L381--L384

\bibitem[Reach {\it et al.}(1997)]{reachband} \refic Reach, W. T., B. A. Franz, 
J. L. Weiland;1997;The Three-Dimensional Structure of the Zodiacal Dust 
Bands;Icarus;127;461--484

\bibitem[Reach, Wall, and Odegard(1998)]{reach98} \refic Reach, W. T.,
W. F. Wall, and N. Odegard;1998;Infrared excess and molecular clouds:
a comparison of new surveys of far-infrared and H~I 21 centimeter
emission at high galactic latitudes;\apj;507;507--525

\bibitem[Reach {\it et al.}(2000)]{reachencke} \refic Reach, W. T., M. 
V. Sykes, D. Lien, and J. K. Davies;2000;The Formation of Encke 
Meteoroids and Dust Trail;Icarus;148;80--94

\bibitem[Renard, Levasseur-Regourd, and Dumont(1995)]{renard} \refic Renard, J. B.,
A. C. Levasseur-Regourd, and R. Dumont;1995;Properties of interplanetary
dust from infrared and optical observations. II. Brightness,
polarization, temperature, albedo and their dependence on elevation
above the ecliptic;\aap;304;602--609

\bibitem[Sandford(1987)]{sandford} \refic Sandford, S. A.;1987;The collection and
analysis of extraterrestrial dust particles;Fund. Cosmic Phys.;12;1--73

\bibitem[Sandford and Walker(1985)]{sandfordwalker} \refic Sandford, 
S. A. and R. M. Walker;1985;Laboratory infrared transmission spectra of 
individual interplanetary dust particles from 2.5 to 25 
microns;\apj;291;838--851

\bibitem[Scott and Duley(1996)]{scott96}  \refic Scott, A.~and W.~W.
Duley;1996;Ultraviolet and Infrared Refractive Indices of 
Amorphous Silicates;\apjs;105;401--405

\def\extra{
\bibitem[van de Hulst(1957)]{1957lssp.book.....V} {\noindent \pp{\sc van de Hulst, H. 
C.} {Light Scattering by Small Particles}. 1957. (John Wiley \& Sons: New York)}
}

\bibitem[Telesco et al.(1991)]{telknack}
\refic Telesco, C. M., and R. F. Knacke;1991;Detection of 
silicates in the Beta Pictoris disk;\apj;372;L29

\bibitem[Tran {\it et al.}(2001)]{tran01} \refic Tran, Q. D. and 12 
colleagues;2001;ISOCAM-CVF 5--12 micron spectroscopy of ultraluminous
infrared galaxies;\apj;552;527--543

\bibitem[van den Ancker {\it et al.}(2001)]{vandenancker} \refic van den Ancker, M. E.,
G. Meeus, J. Cami, L. B. F. M. Waters, and C. Waelkens;2001;The
composition of circumstellar gas and dust in 51 Oph;\aap;369;L17--L21

\bibitem[Waelkens {\it et al.}(1996)]{waelkens} \refic Waelkens, C., and 20 
coauthors;1996;SWS observations of young main-sequence stars with dusty 
circumstellar disks;\aap;315;L245--L248

\bibitem[Weinberger {\it et al.}(2002)]{weinberger} \refic Weinberger, A. J.,
E. E. Becklin, and B. Zuckerman;2002;Spatially resolved infrared imaging
and spectroscopy of $\beta$ Pictoris;\apj;;{in preparation}

\bibitem[Whipple(1955)]{whipple} \refic Whipple, F. L.;1955;A  comet model. III. The zodiacal 
light;\apj;121;750--770

\bibitem[Wooden {\it et al.}(1999)]{wooden} \refic Wooden, D. H., D. E. Harker, C. E. Woodward,
H. M. Butner, C. Koike, F. C. Witteborn, and C. W. McCurtry;1999;Silicate mineralogy
of the dust in the inner coma of comet C/1995 O1 (Hale-Bopp)
pre- and post-perihelion;\apj;517;1034--1058

\bibitem[Wright(1998)]{wright98} \refic Wright, E. L.;1998;Angular Power Spectra of the 
COBE DIRBE Maps;\apj;496;1--8

\bibitem[Yanamandra-Fisher and Hanner(1999)]{yanamandra} \refic Yanamandra-Fisher, P. A.,
and M. S. Hanner;1999;Optical Properties of Nonspherical Particles of Size Comparable to the 
Wavelength of Light: Application to Comet Dust;Icarus;138;107--128

\end{thebibliography}
\end{document}